\documentclass[tradiabstract]{aa} 
\usepackage{graphicx}
\usepackage{natbib}
\usepackage{subfigure}
\begin{document}
   \title{Frequency and properties of bars in cluster and field
     galaxies at intermediate redshifts\thanks{Based on
     observations collected at the European Southern Observatory,
     Chile, as part of large programme 166.A-0162 (the ESO Distant
     Cluster Survey). Also based on observations made with the
     NASA/ESA $Hubble$ $Space$ $Telescope$, obtained at the Space
     Telescope Science Institute, which is operated by the Association
     of Universities for Research in Astronomy, Inc., under NASA
     contract NAS 5-26555. These observations are associated with
     proposal 9476. Support for this porposal was provided by NASA
     through a grant from Space Telescope Science Institute.}}


   \author{Fabio D. Barazza\inst{1} \and Pascale Jablonka\inst{1,2,3}
     \and Vandana Desai\inst{4} \and Shardha Jogee\inst{5} \and Alfonso
     Arag\'on-Salamanca\inst{6} \and Gabriella De Lucia\inst{7} \and
     Roberto P. Saglia\inst{8} \and Claire Halliday\inst{9} \and
     Bianca M. Poggianti\inst{10} \and Julianne J. Dalcanton \inst{11} \and
     Gregory Rudnick\inst{12} \and Bo Milvang-Jensen\inst{13,14} \and
     Stefan Noll\inst{8,15} \and Luc Simard\inst{16} \and 
     Douglas I. Clowe\inst{17} \and Roser Pell\'o\inst{18} \and Simon
     D.M. White\inst{7} \and Dennis Zaritsky\inst{19}
          }

   \institute{Laboratoire d'Astrophysique, \'Ecole
Polytechnique F\'ed\'erale de Lausanne (EPFL), Observatoire de
Sauverny CH-1290 Versoix, Switzerland \email{fabio.barazza@epfl.ch}
\and
Universit\'e de Gen\`eve, Observatoire de Sauverny CH-1290 Versoix,
Switzerland \and
GEPI, CNRS-UMR8111, Observatoire de Paris, section de Meudon, 5 Place
Jules Janssen, F-92195 Meudon Cedex, France \and
{\it Spitzer} Science Center, Caltech, Pasadena CA 91125, USA \and
Department of Astronomy, University of Texas at Austin, 1 University
Station C1400, Austin, TX 78712-0259, USA \and
School of Physics and Astronomy, University of Nottingham, University
Park, Nottingham NG7 2RD, UK \and
Max-Planck-Institut f\"ur Astrophysik, Karl-Schwarzschild-Strasse 1,
D-85748 Garching bei M\"unchen, Germany \and
Max-Planck-Institut f\"ur extraterrestrische Physik,
Giessenbachstrasse, D-85748 Garching bei M\"unchen, Germany \and
INAF, Osservatorio Astrofisico di Arcetri, Largo E. Fermi 5, I-50125
Firenze, Italy \and
Osservatorio Astronomico di Padova, Vicolo dell'Osservatorio 5, 35122
Padova, Italy \and
Astronomy Department, University of Washington, Box 351580, Seattle,
WA 98195, USA \and
NOAO, 950 N. Cherry Avenue, Tucson, AZ 85719, USA \and
Dark Cosmology Centre, Niels Bohr Institute, University of Copenhagen, 
Juliane Maries Vej 30, DK-2100 Copenhagen, Denmark \and
The Royal Library / Copenhagen University Library, Research Dept., Box 
2149, 1016 Copenhagen K, Denmark \and
Observatoire Astronomique de Marseille-Provence, 38 rue Fr\'ed\'eric
Joliot-Curie, 13388 Marseille cedex 13, France \and
Herzberg Institute of Astrophysics, National Research Council of
Canada, 5071 West Saanich Road, Victoria, Canada BC V9E 2E7 \and
Ohio University, Department of Physics and Astronomy, Clippinger Labs
251B, Athens, OH 45701, USA \and
Laboratoire d'Astrophysique de Toulouse-Tarbes, CNRS, Universit\'e de
Toulouse, 14 Avenue Edouard Belin, F-31400 Toulouse, France \and
Steward Observatory, University of Arizona, 933 North Cherry Avenue,
Tucson, AZ 85721, USA
             }

   \date{Received; accepted}

 
  \abstract
   {We present a study of large-scale bars in field and cluster
     environments out to redshifts of $\sim0.8$ using a final
     sample of 945 moderately inclined disk galaxies drawn from the
     EDisCS project. We characterize  bars and their host galaxies and
     look for relations between the presence of a bar and the properties
     of the underlying disk. We investigate whether the fraction and
     properties of bars in clusters are different from their
     counterparts in the field. The properties of bars and disks are
     determined by ellipse fits to the surface brightness distribution
     of the galaxies using HST/ACS images in the F814W filter. The bar
     identification is based on quantitative criteria after highly
     inclined ($>60^{\circ}$) systems  have been excluded. The total
     optical bar fraction in the redshift range $z=0.4-0.8$ (median
     $z=0.60$), averaged over the entire sample, is $25\%$ ($20\%$ for
     strong bars). For the cluster and field subsamples, we
     measure bar fractions of $24\%$ and $29\%$, respectively. We find
     that bars in clusters are on average longer than in the field and
     preferentially found close to the cluster center, where the bar
     fraction is somewhat higher ($\sim31\%$) than at larger distances
     ($\sim18\%$). These findings however rely on a relatively
       small subsample and might be affected by small number
       statistics. In agreement with local studies, we find
     that disk-dominated galaxies have a higher optical bar fraction
     ($\sim45\%$) than bulge-dominated galaxies ($\sim15\%$). This
     result is based on Hubble types and effective radii and does not
     change with redshift. The latter finding implies that bar
     formation or dissolution is strongly connected to the emergence
     of the morphological structure of a disk and is typically
     accompanied by a transition in the Hubble type. The question
     whether internal or external factors are more important for bar
     formation and evolution cannot be answered definitely. On the one
     hand, the bar fraction and properties of cluster and field
     samples of disk galaxies are quite similar, indicating that
     internal processes are crucial for bar formation. On the other
     hand, we find evidence that cluster centers are favorable
     locations for bars, which suggests that the internal processes
     responsible for bar growth are supported by the typical
     interactions taking place in such environments.

   \keywords{}}
   \titlerunning{Bars in cluster and field galaxies at intermediate
     redshifts}
   \maketitle
%

\section{Introduction}
There is evidence that the dynamical and secular evolution
of disk galaxies is intimately connected with the presence of stellar
bars. Theory and $n$-body simulations predict that bars transfer
angular momentum to the outer disk, which causes the stellar orbits in
the bar to become elongated and the bar amplitude to increase
\citep{lyn79,pfe91,sel93,ath03}. The growing bar becomes more and more 
efficient in driving gas inside the corotation radius towards the
center of the disk, which can trigger starbursts
\citep{hun99,reg99,sak99,reg04,bou02,sch02,jog05,sht05} and
contribute to the formation of disky bulges
\citep{kor93,sak99,kor04,ath05,jog05,sht05,deb06}. The
redistribution of angular momentum driven by bars is not restricted to
the baryonic component, but also applies to dark matter
\citep{wei85,com90,deb00,ber06}. Another indication of secular
evolution induced by the orbital structure and resonances in a bar
potential is provided by box- or peanut-shaped bulges in inclined
galaxies \citep{com90,pfe90,kui95,bur99,mar06,deb06}. These processes
affect the properties of disk galaxies and therefore shape
the diversity of morphologies.

While it is still unknown why a specific disk galaxy hosts a bar and
an apparently similar galaxy is unbarred, it is clear that a
significant fraction of bright disk galaxies appears barred in optical
observations \citep{esk00,mar07,ree07,bar08}. In studies carried out in the
near-infrared (NIR) or in both, NIR and optical, the NIR bar fractions
are typically higher \citep{kna99,esk00,lau04,men07,mar07}. These
findings indicate that bar detection is affected by dust extinction
and that studies completed at different wavelengths cannot be readily
compared. This caveat is important particularly when results
from local bar studies are compared with the findings of investigations
at intermediate redshifts, where issues such as band shifting, surface
brightness dimming, and reduced resolution further decrease the bar
detection rate. In earlier studies, it was found that the bar
fraction undergoes a significant intrinsic decline out to $z\sim1$
\citep{abr99,van00}, which was confirmed by
\cite{sht08}. Other studies report that the bar fraction is
fairly constant out to $z\sim1$ for strong bars \citep[bar ellipticity
  $>0.4$]{jog04} or all bars \citep[][see also Sect.
  \ref{discu}]{elm04,zhe05}, once the aforementioned effects are taken
into account.

To identify and characterize bars, different methods have been 
applied. The most straightforward approach is to inspect images
visually and assign a bar class (e.g., weak/strong bars) to each galaxy
\citep{dev91,esk00}. By adopting this approach, strong bars were
  detected twice as frequently in near-infrared data than optical data
\citep{esk00} and the bar fraction was found to increase between Sc galaxies
and later types \citep{ode96}. Apart from visual classification, a
quantitative characterization of bars has been attempted using for
instance, the gravitational torque method \citep{blo02,lau02,but05},
Fourier dissection \citep{but06,lau06}, and ellipse fits to the galaxy
isophotes
\citep{reg97,abr99,sht00,sht02,sht08,kna00,erw02,erw05,jog02a,jog02b,jog04,why02,elm04,ree07,men07,mar07,bar08}. These
methods provide measurements of the bar length and shape and
enable the impact of the bar on the disk to be assessed.

The vast majority of these bar studies have concentrated on field
galaxy samples and estimated the bar fraction
among disk galaxies and the general properties of bars and their host
galaxies. First attempts have been made to relate the presence of a bar to
its host galaxy properties, such as disk structure, central surface
brightness, or color. On the other hand, there have been few studies
of the relation between the occurrence of bars and
environment. \citet{kum86} studied the relative fractions of different
disk galaxy types as a function of environment. They detected no
increase in bar fraction inside galaxy clusters or groups, but
measured a significantly higher fraction of barred disks in binary
systems. This was confirmed by \citet{elm90} and \citet{giu93}, who
both found that galaxies in binary systems are preferentially early
type and barred. A similar result was reported by
\citet{var04}. Interestingly, while the fraction of barred disks in
clusters or groups is not higher than in the field
\citep{kum86,van02,van07,mar09}, \citet{tho81} and \citet{and96} presented
evidence that barred galaxies in the Coma and Virgo clusters are more
concentrated toward the cluster centers than unbarred disks.

We present the first study of bars in clusters at
intermediate redshifts, which enables the properties of bars and their host
galaxies to be studied in dense environments. We use a final sample of
945 moderately inclined disk galaxies drawn from a parent
  sample of 1906 disk galaxies from the ESO  distant cluster survey
\citep[EDisCS,][]{whi05}. We use the
available $I$-band $HST/ACS$ images to identify and characterize bars,
based on quantitative criteria. We use this sample to look for
  relations between the occurrence and the properties of bars and
  their host galaxies. For a subsample of 241
  objects, for which spectroscopic redshifts and reliable cluster
  membership determinations are available, we look for relations
  between barred and unbarred galaxies and their environment. We also
  study a wide range of redshifts, which allows us to search for
  trends with increasing look-back time. The
outline of the paper is as follows: In Sect., \ref{samsel} we present
the available data and our sample selection. The method to identify and 
characterize bars as well as limits of the detectability of bars in
our data is described in Sect. \ref{charac}. Our results in terms of
bar fractions and relations between bars and host galaxy properties
are presented in Sect. \ref{resul}. In Sect., \ref{bprop} we discuss the
properties of the bars in our sample galaxies, and in Sect. \ref{cldis}
look at the specific distribution of barred galaxies within the
clusters. The implications of our results are discussed in Sect.
\ref{discu} and the summary and conclusions are given in Sect.
\ref{sum}. Throughout the paper, we assume a flat cosmology with
$\Omega_M=1-\Omega_{\Lambda}=0.3$ and $H_0=70$ km~s$^{-1}$
Mpc$^{-1}$. Magnitudes are given in the Vega system.

\section{Sample selection from EDisCS}\label{samsel}
The ESO Distant Cluster Survey is a study of 26 optically selected and
spectroscopically confirmed galaxy systems, from rich groups to
massive clusters, smoothly distributed between redshifts 0.39 and 0.96 
\citep{hal04,mil08}. For all systems, we have assembled three-band optical VLT
deep photometry, deep NTT/SOFI near-infrared imaging, and optical
VLT/FORS2 spectroscopy \citep{whi05}. We also acquired HST/ACS images
in the $F814W$ filter for 10 fields containing the most distant
clusters studied by EDisCS. The exposure time of these
  observations per pixel is 2040 s, except for the central $3.5\arcmin
  \times 3.5\arcmin$, which has an exposure time per pixel of 10,200 s.
We completed both a visual classification
of galaxy morphologies (see Sect. \ref{visclas}) and a quantitative
analysis of their structural parameters  
\citep{des07,sim09}. The visual classification was completed for all
galaxies brighter than $I_{auto}=23$ mag, where $I_{auto}$ is the
SExtractor AUTO magnitude measured on the $I$-band VLT images. In this
work, we adopt the same magnitude cut at $I_{auto}=23$ mag and
consider only galaxies with Hubble types S0--Sm/Im, i.e. disk galaxies and
lenticulars, which can also be barred \footnote{In the remainder of
  the paper, we use the terms disk--galaxy sample or disk galaxies for
  brevity including S0s.}. For $\sim90\%$ of the sample, a
S\'ersic fit to the surface brightness distribution \citep[with
GIM2D,][]{sim02,sim09} has been performed on the $I$-band $HST/ACS$
images, providing the effective radius used in our analysis (see
Sect. \ref{seff}).
\begin{table*}
\caption{Basic properties of clusters and the secondary structures}
\label{basic}
\centering
\begin{tabular}{l c c c r r r r}
\hline\hline
\multicolumn{1}{c}{Cluster/Group} & R.A. (J2000.0) & Decl. (J2000.0) & $z$ &
\multicolumn{1}{c}{$\sigma$} & \multicolumn{1}{c}{$R_{200}$} &
\multicolumn{1}{c}{$N_{tot}^C$} & \multicolumn{1}{c}{$N_{tot}^F$} \\
\multicolumn{1}{c}{Name} & (hh mm ss) & (dd mm ss) & & \multicolumn{1}{c}{(km s$^{-1}$)} &
\multicolumn{1}{c}{(Mpc)} & & \\
\hline
cl1037.9-1243 & 10 37 51.4 & -12 43 26.6 & 0.58 & 319 & 0.57 & 168 & 74\\
cl1037.9-1243a & 10 37 52.3 & -12 44 49.0 & 0.43 & 537 & 1.06 & 33 & (...)\\
cl1040.7-1155 & 10 40 40.3 & -11 56 04.2 & 0.70 & 418 & 0.70 & 68 & 86\\
cl1054.4-1146 & 10 54 43.5 & -11 46 19.4 & 0.70 & 589 & 0.99 & 138 & 74\\
cl1054.7-1245 & 10 54 43.5 & -12 45 51.9 & 0.75 & 504 & 0.82 & 94 & 89\\
cl1103.7-1245 & 11 03 43.4 & -12 45 34.1 & 0.96 & 534 & 0.77 & 0 & 76\\
cl1103.7-1245a & 11 03 34.9 & -12 46 46.2 & 0.63 & 336 & 0.59 & 3 & (...)\\
cl1103.7-1245b & 11 03 36.5 & -12 44 22.3 & 0.70 & 252 & 0.42 & 6 & (...)\\
cl1138.2-1133 & 11 38 10.2 & -11 33 37.9 & 0.48 & 732 & 1.40 & 103 & 72\\
cl1138.2-1133a & 11 38 08.6 & -11 36 54.9 & 0.45 & 542 & 1.05 & 7 & (...)\\
cl1216.8-1201 & 12 16 45.3 & -12 01 17.6 & 0.79 & 1018 & 1.61 & 117 & 127\\
cl1227.9-1138 & 12 27 53.9 & -11 38 17.3 & 0.64 & 574 & 1.00 & 128 & 76\\
cl1227.9-1138a & 12 27 52.1 & -11 39 58.7 & 0.58 & 341 & 0.61 & 8 & (...)\\
cl1232.5-1250 & 12 32 30.3 & -12 50 36.4 & 0.54 & 1080 & 1.99 & 143 & 66\\
cl1354.2-1230 & 13 54 09.8 & -12 31 01.5 & 0.76 & 648 & 1.05 & 60 & 81\\
cl1354.2-1230a & 13 54 11.4 & -12 30 45.2 & 0.60 & 433 & 0.77 & 7 & (...)\\
\hline
\\
\end{tabular}
\begin{minipage}{\textwidth}
Notes: When more than one cluster or group is found in the EDisCS
fields, they are identified with 'a' or 'b' appended to the name of
the main cluster (structures with $\sigma<400$ kms$^{-1}$ are regarded 
as groups). The coordinates of the clusters/groups correspond to
the positions of the brightest cluster/group galaxies. The redshifts
($z$) and line-of-sight cluster/group velocity dispersions ($\sigma$)
are taken from \citet{hal04} and \citet{mil08}. The virial radius
($R_{\rm 200}$) has been determined using eq. (8) from
\citet{fin05}. $N_{tot}^C$ gives the total number of {\it disk} galaxies
(i.e. S0--Sm/Im) in the cluster and $N_{tot}^F$ gives the number of
{\it disk} galaxies in the corresponding field. For the structures
cl1227.9-1138 and cl1227.9-1138a no S\'ersic fits could be performed,
because the data only became available much later and could not be
included in the general analysis anymore.
\end{minipage}
\end{table*}
\begin{figure}
\centering
\includegraphics[width=9cm]{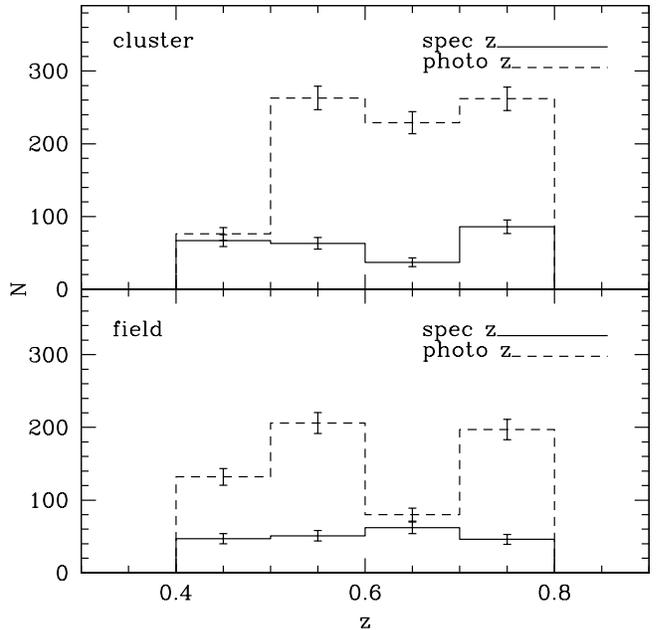}
\caption{Redshift distribution of the basic sample of 1906 disk
  galaxies. The solid line shows the distribution based on
  spectroscopic redshifts and the dashed line that based on
  photometric redshifts.}
\label{reds}
\end{figure}
\begin{figure}
\centering
\includegraphics[width=9cm]{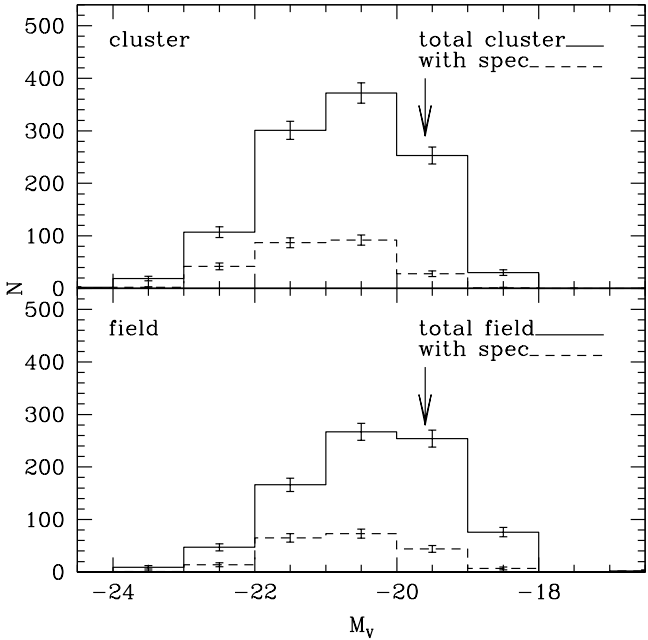}
\caption{Distribution of absolute $V$ magnitude for the cluster and
  field samples. The solid histograms show the distribution for the
  total sample of 1906 galaxies, whereas the dashed histograms
  represent the spectroscopic subsample. The arrows indicate the
  magnitude cut (at $M_V\approx-19.6$ mag) for a Sa galaxy at $z=0.8$
  corresponding to  $I_{auto}=23$ mag.}
\label{vmags}
\end{figure}

We select all galaxies meeting these criteria in ten fields regardless
of whether they are cluster members or group/field galaxies. These
fields encompass a total of 12 clusters and 4 groups.
  Structures with $\sigma<400$ kms$^{-1}$ are regarded as
  groups. Galaxies are considered to be cluster/group members if the
  integrated photometric redshift probability to be within $z\pm0.1$
  of the cluster redshift is greater than a specific limit. These
  limits are based on our spectroscopy and range from 0.1 to 0.35
  depending on the filter set available for each particular field
  \citep{whi05,pel08}. Cluster/group membership for objects with
  spectroscopic observations is defined as being within
  $\pm3\sigma_{cl}$ of $z_{cl}$ \citep{hal04,mil08}. Photometric
redshifts were determined using the methods described in \cite{rud01}
and \cite{pel08},
and were based on the optical+near-infrared photometry. The
  accuracy of photometric redshifts is typically $\Delta
  z/(1+z)\approx0.05\pm0.01$ \citep{pel08}. The rest-frame
magnitudes and colors were computed using the method described
in \citet{rud03}. We restrict the cluster and field samples to
  the redshift range $z=0.4-0.8$ in order to remain in the rest-frame
  optical. The median photometric redshift of the total sample is
0.60. The basic properties of the main and secondary structures and
the number of objects found in these structures are given in Table
\ref{basic}. We show the distributions in redshift and absolute $V$
magnitude in Figs. \ref{reds} and \ref{vmags}, respectively. These
plots include the parent sample of 1906 galaxies, to
which our bar classification method (see Sect. \ref{charac}) was
applied. Spectroscopic redshifts are available for a subsample of 459
galaxies. For the low redshift cluster fields
($z<0.6$) spectroscopic observations were restricted to objects
with $I<22$ mag and for the high redshift cluster fields ($z>0.6$) to
$I<23$ mag \citep{hal04,mil08}. This ensures that the distribution in
absolute magnitude for galaxies with spectroscopic redshifts remains
roughly the same over the entire redshift range ($z=0.4-0.8$). The
distributions in Figs. \ref{reds} and \ref{vmags} for the cluster
and field subsamples are very
similar. The arrows in Fig. \ref{vmags} indicate the absolute
$V$-band magnitude for a Sa galaxy at $z=0.8$ corresponding to
$I_{auto}=23$ mag ($M_V\approx-19.6$ mag, taking into account a
K-correction). The same calculation for a Sc galaxy would result in a slightly
fainter absolute $V$-band magnitude. In the following, we present
  our main results for both the total sample and a sample restricted to
  $M_V\leq-20$ mag, which is our completeness limit  (the complete sample). This
  shows that the incompleteness for galaxies with $M_V>-20$ is not biasing our
  results. All numbers and fractions always refer to the total sample. In Sect.
\ref{resul}, results based on the separation between cluster/group and
field galaxies are always based on our spectroscopic data. This
reduces the sample size considerably, but ensures a
reliable cluster or field allocation. We estimate that a
photometrically based cluster sample would have a field galaxy
contamination of up to $40\%$. Finally, we emphasize that the number
of objects in the four groups is rather small and that only three bars are
found in group galaxies (see also Table \ref{basic}). For the
remainder of the paper, we therefore refer to the subsample of
cluster/group galaxies as the cluster subsample.

\subsection{The visual classification of the galaxies}\label{visclas}
We visually classified all galaxies brighter than $I_{auto}=23$ mag,
where $I_{auto}$ is the SExtactor \citep{ber96} magnitude measured on
the $I$-band VLT images and is an estimate of the total magnitude of a
galaxy in the Vega system. This classification has been described in
detail in \cite{des07}. We recall here its main features in particular
those related to our purpose, i.e., the spiral galaxies. Each of the
five classifiers (AAS, JJD, VD, PJ and BP ) was trained on the HST
WFPC2 images and visual morphological catalogs of the $0.3<z<0.5$
MORPHS clusters, using the same procedure as described in \cite{sma97}.

\begin{figure*}
\centering
\includegraphics[width=15cm]{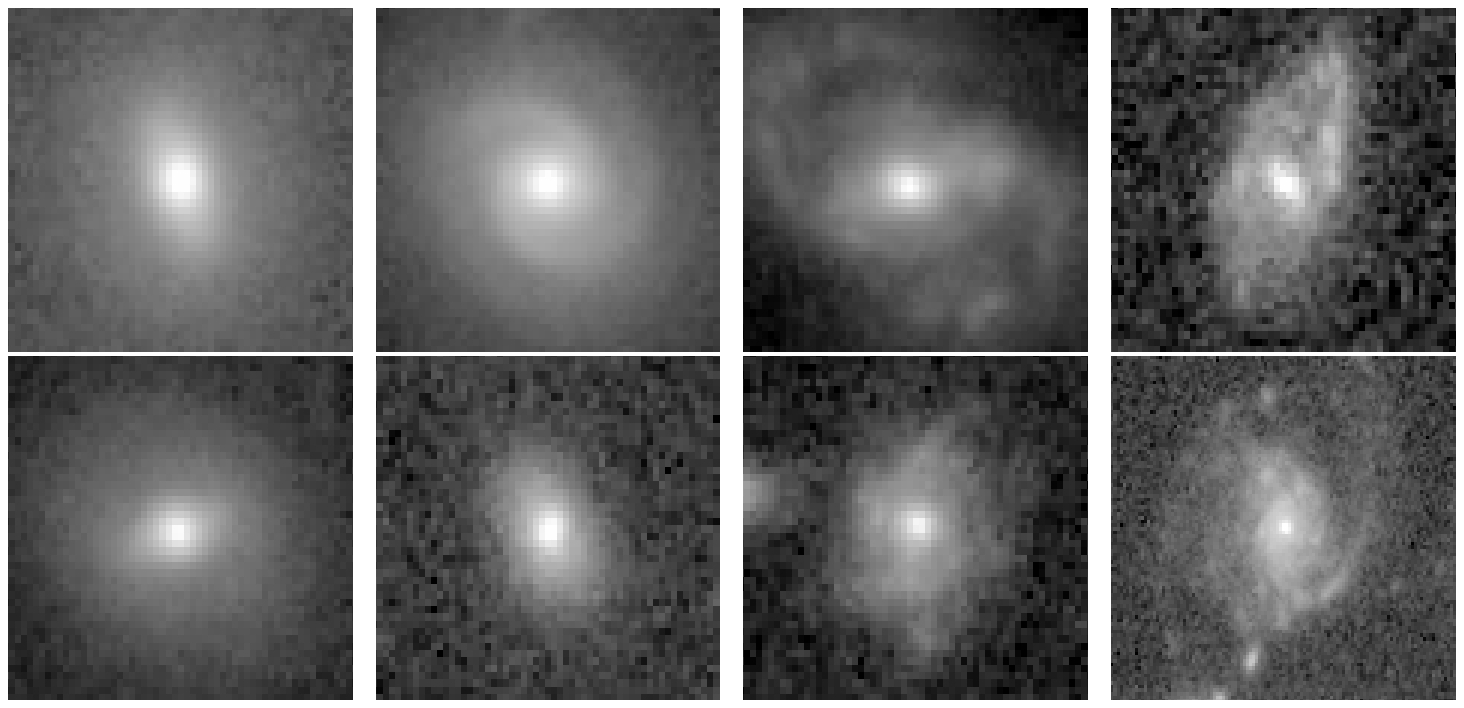}
\caption{From left to right examples of S0, Sa, Sb, and Sc galaxies
  for cluster CL1232-1250 at $z=0.54$ (top row) and cluster
  CL1216-1201 at $z=0.79$ (bottom row). The images are
  $\sim2\farcs5\times2\farcs5$.}
\label{mclass}
\end{figure*}

Since, in this analysis, we consider possible trends of bar fraction
with the Hubble type of the galaxies, we now investigate our ability
to distinguish between adjacent types, e.g., Sa from Sb, Sc from Sd, etc.
Galaxies in the cluster Cl1216-1201 ($z=0.79$) were classified by all
five classifiers and the data set for this cluster can be used to
complete the most reliable statistical analysis. We found that the mean
dispersion between classifiers and for all spiral galaxies was 1.2
T-type. The fraction of galaxies with a dispersion of less than
2 T-types among galaxies was 66\%, and increased to 89\% for 3-Types
(Sa to Sb for example). These global numbers did not change if one
considers early- or late-type spirals. Type 6 (Sd) galaxies were
identified within 3 T-types at 88\%, while Type 1 (Sa) were at
91\%. The rest of the clusters were analyzed by two classifiers. Their
statistics is either identical to the one of Cl1216-1201 or have
slightly lower success rates. However, the ability to agree to less
than 3 T-type is never lower than 60\%. This shows that our
classification is robust and trends along the Hubble sequence can
be reliably detected. In Fig. \ref{mclass} we show examples of
S0--Sc galaxies for cluster CL1232-1250 at $z=0.54$ and cluster
CL1216-1201 at $z=0.79$. The image depth and resolution are sufficient
to separate Hubble types even at $z=0.79$.

\section{Bar characterization and detectability}\label{charac}
We first describe our method to detect and characterize bars using the 
$I$-band $HST/ACS$ images and then discuss the limitations of the
detectability of bars imposed by the observations used in our study.

\subsection{The detection and characterization of bars}\label{method}
Our method of finding bars relies on the fact that the isophotes of bars
in moderately inclined disk galaxies (i.e., with disk inclination
$i<60^{\circ}$) have much higher ellipticities than the isophotes of
the underlying disk. The ellipticities of the isophotes are derived
by fitting ellipses to the surface brightness distribution of the
disks. The corresponding profiles of ellipticity ($\epsilon$) and
position angle (P.A.) are investigated based on quantitative
criteria. The method of ellipse fits has been used widely by
observational studies of bars in disk galaxies
\citep{fri96,reg97,abr99,jog99,kna00,sht00,lai02,why02,jog02a,jog02b,sht03,elm04,ree07,men07,sht08}.
There is also strong theoretical evidence supporting this approach 
\citep{ath92,she04}.

The specific method we use in this study has already been applied in
earlier investigations and a detailed description can be found in the
corresponding papers \citep{jog04,mar07,bar08}. We start by fitting
ellipses to the images using the standard IRAF task `ellipse' via an
iterative wrapper developed by \cite{jog04}, which (a) refines the
center of the galaxy using the IRAF routine `imcenter'; (b) determines
the maximum galaxy semi-major axis length ($a_{\rm max}$) out to which
ellipses will be fitted by finding where the galaxy isophotes
reach the sky level; (c) executes `ellipse' for a maximum number $N$
of iterations, for each object, analyzing each output on the fly to
guide the next fit. The wrapper stops either if all the isophotes can
be fitted or if the maximum $N$ of iterations is reached, where $N$ is
typically set to be 300. A fit is considered to be
successful if all the isophotes can be
fitted either in one single fit or via a combination of partial
fits. The fitting process for an individual galaxy can fail completely 
if the center of an object cannot be found or if the surface
brightness oscillates too strongly across the galaxy. This typically
happens when a foreground star is present or when two objects
overlap. For only $\sim7\%$ of objects in our initial sample, the
ellipse fit failed and successful fits were obtained for 1906
galaxies. When using the IRAF task `ellipse' for ellipse fits, the
goodness of the best-fit solution is measured by four harmonic
amplitudes (A3, A4, B3, B4), which describe by how much the true
isophote differs from the best-fit model ellipse
\citep[e.g.,][]{jed87}. We find that the  deviations from ellipses are
typically small ($<10\%$).

\begin{figure*}
\centering
\includegraphics{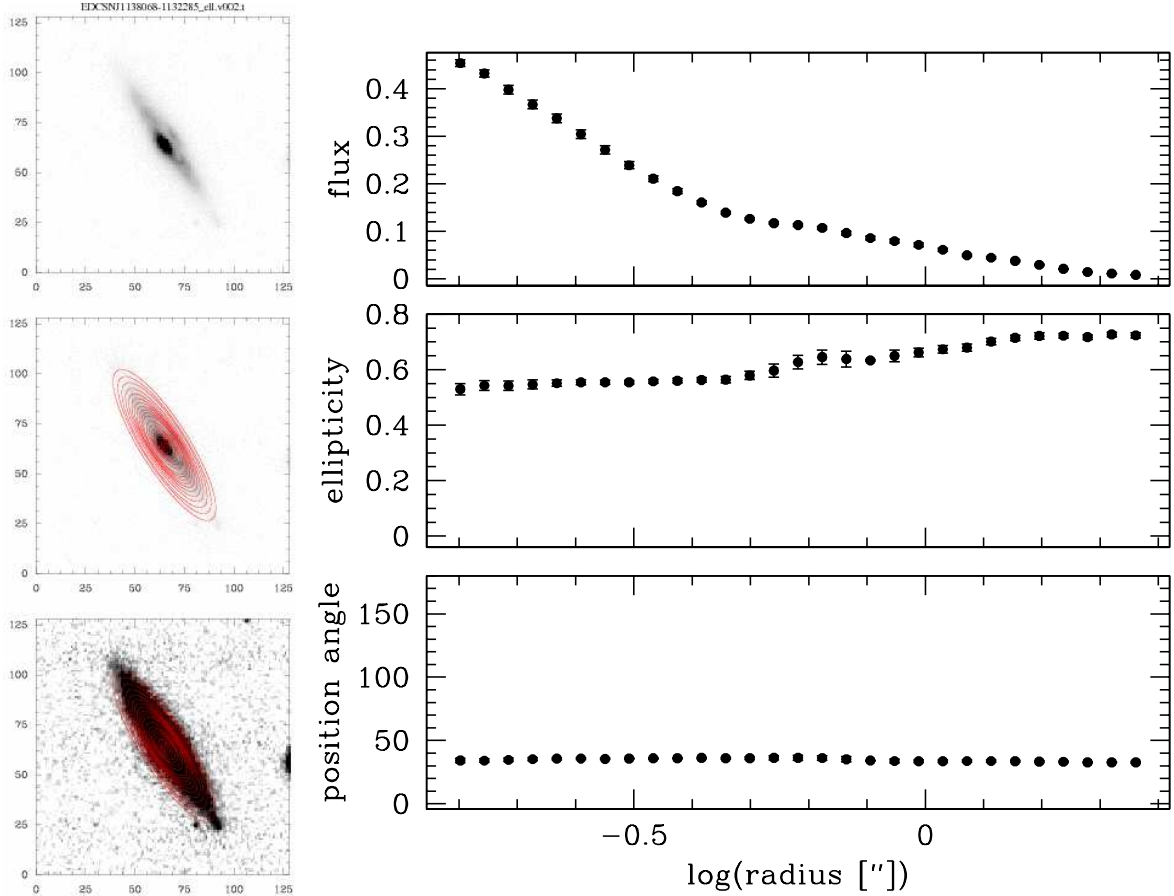}
\caption{This is an example of an inclined ($i>60^{\circ}$) galaxy,
  identified from the overlays and radial profiles that are generated by the
  ellipse fits. Such galaxies are excluded from our final sample. {\it
    Left:} The top image shows only the galaxy, while the middle and
  bottom images show the ellipses overlaid on the galaxy, with
  different greyscale stretches to emphasize the inner and outer
  regions of the galaxy. The images are roughly $6\arcsec$ on a
  side. {\it Right:} The radial profiles of surface brightness (top),
  ellipticity $\epsilon$ (middle), and PA (bottom) are shown. In the
  outer parts of the galaxy, the PA is flat and the ellipticity is
  fairly constant at $\epsilon >0.5$, showing that the galaxy has a
  high inclination ($i>60^{\circ}$).}
\label{inclined}
\end{figure*}

\begin{figure*}
\centering
\includegraphics{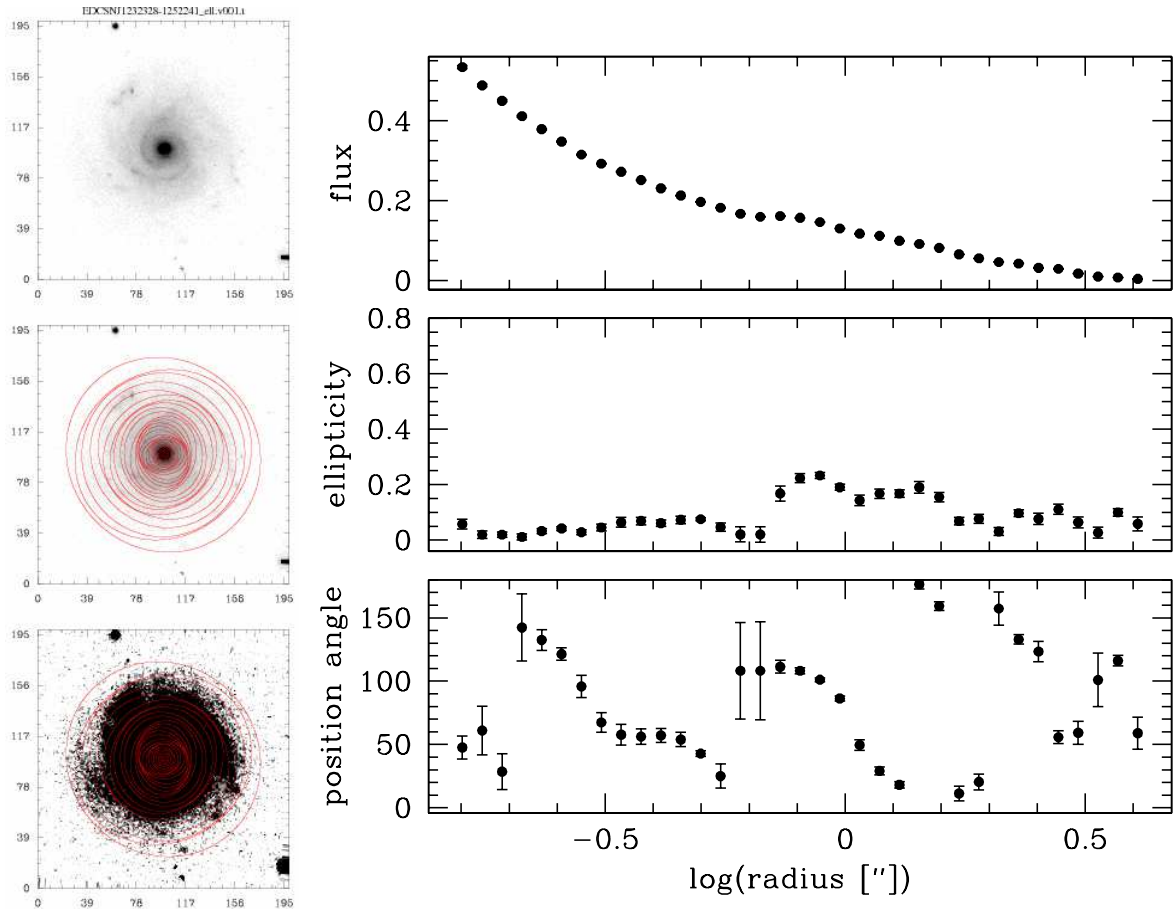}
\caption{The same as Fig. \ref{inclined}, but for a galaxy
  classified as unbarred according to our ellipse fits. The images are 
  roughly $10\arcsec$ on a side.}
\label{unbarred}
\end{figure*}

\begin{figure*}
\centering
\includegraphics{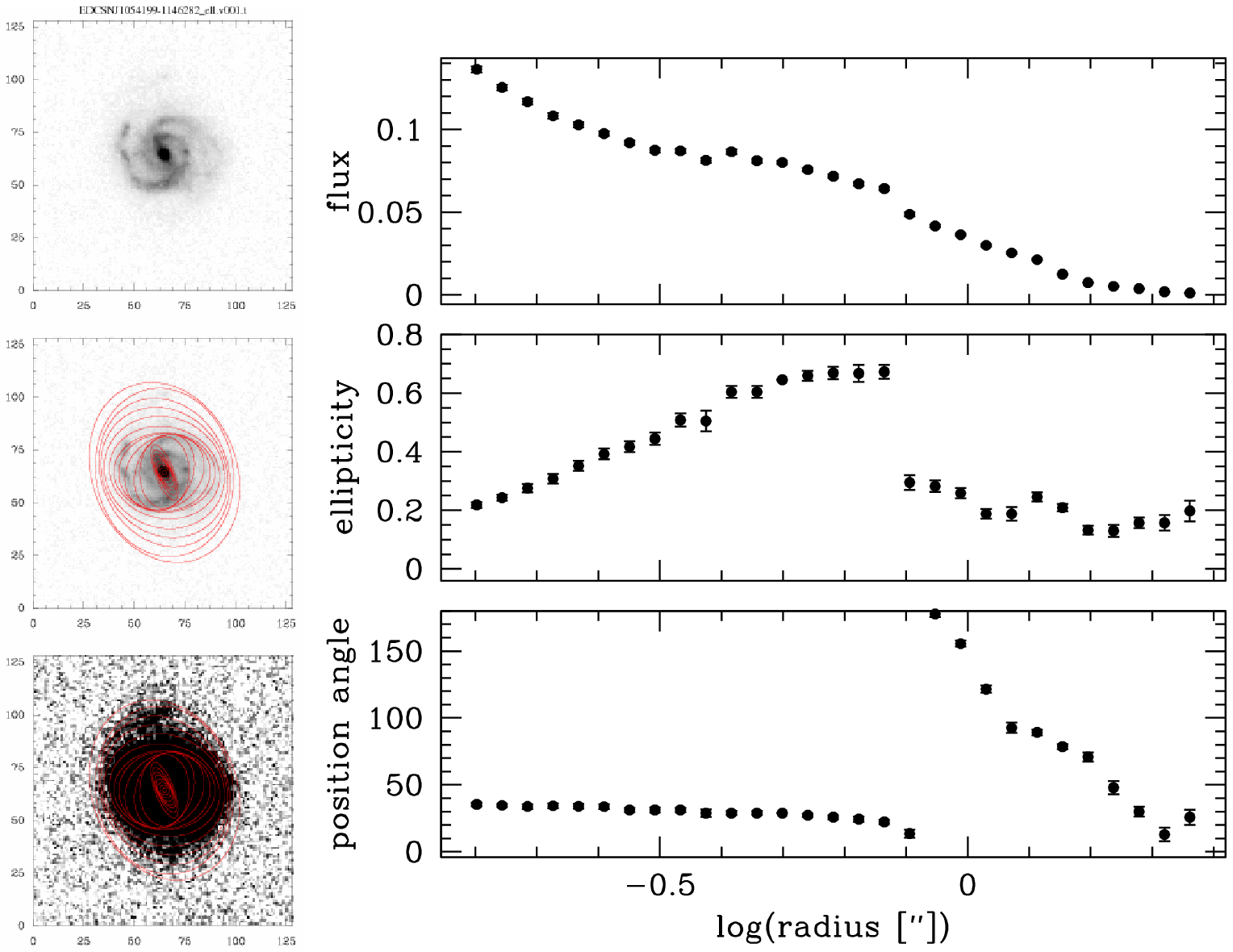}
\caption{The same as Fig. \ref{inclined}, but for a galaxy
  classified as barred from ellipse fits. The images are roughly
  $6\arcsec$ on a side. Over the bar region, $\epsilon$ rises smoothly
  to a global maximum of $\sim0.7$, while the PA remains approximately
  constant. After the end of the bar at the transition to the more circular
  disk, the ellipticity decreases sharply at $\sim0\farcs8$ and the PA
  starts to change significantly at this point.}
\label{barred}
\end{figure*}

Based on these fits, radial profiles of the surface brightness,
$\epsilon$, and P.A. are derived, and the fitted ellipses are
overplotted onto the galaxy images. Examples of these profiles and
overlays are shown in Figs. \ref{inclined}, \ref{unbarred}, and
\ref{barred}. These representations and the true images
are the primary tools for the classification. In a first step, the disk
inclination ($i$) is determined using the $\epsilon$ profile. For each
galaxy, an interactive visualization tool \citep{jog04} is used to
display the overlay and radial profiles, interactively determine the
disk parameters (ellipticity, PA) and bar parameters (maximum
ellipticity, semi-major axis, PA), and assign the main classification
of `inclined', `unbarred', and `barred'. If the disk inclination,
i.e., the inclination in the outer parts of the galaxy is higher than
$60^{\circ}$ (or $\epsilon>0.5$) the galaxy is classified as being too
inclined and excluded from further analysis. The galaxy shown in
Fig. \ref{inclined} is such an object. We find that $\sim34\%$ (652
objects) of the sample galaxies have $i>60^{\circ}$, which is similar
to results for other samples \citep{jog04,bar08}, that are comparable
in size and where the same method has been applied.

The remaining galaxies were then classified as unbarred or
barred, based on the following quantitative criteria: (1) $\epsilon$
increases steadily to a global maximum higher than 0.25, while the P.A. 
value remains constant (within $10^{\circ}$), and (2) $\epsilon$ then
drops by at least 0.1 and the P.A. changes at the transition between
the bar to the disk region. Figure \ref{barred} shows a galaxy, which
meets these two criteria. While unbarred spiral galaxies can also
reach large $\epsilon$ and exhibit prominent drops in their $\epsilon$ 
profiles in the region dominated by spiral arms, this is always
accompanied by strong isophotal twists. Since we exclude too inclined
galaxies, criterion (2) is characteristic of a barred disk, because the
disks are typically more circular than the bars for moderately
inclined galaxies. After classifying a galaxy, we used our
interactive display tool to measure the ellipticity, PA, and
semi-major axis of its outer disk. For galaxies classified as barred,
we measure the same quantities, as well as the maximum ellipticity,
$e_{\rm bar}$, of the bar and the semi-major axis, $a_{max}$, of
maximum bar ellipticity. We use $e_{\rm bar}$ as a partial measure of
the bar strength and $a_{max}$ as an estimate for the semi-major axis
of the bar, $a_{bar}$. A detailed theoretical and empirical
justification of this approach is provided in \cite{mar07} and
\cite{men07} (see also the discussion in Sect. \ref{bprop}).

The classifications and quantities measured are based on the observed
images and profiles and are therefore affected by projection effects
(for a detailed discussion of how disk inclinations affect apparent
bar sizes, see \citet{bar08}). We did not attempt to deproject our
galaxies, since it is difficult to determine the P.A. in the outer
disks accurately enough, particularly for galaxies at high
redshifts. The resulting large errors in the P.A. would cause
inaccurate deprojections. We also note that the statistical 
results before and after deprojection are very similar \citep{mar07}.

Of the remaining 1254 moderately inclined ($i<60^{\circ}$) galaxies, we
exclude another 309 objects for two reasons: {\it i)} the
presence of a close neighbor, whose outer isophotes overlap with the
target galaxy, can cause `ellipse' to fit the two galaxies
simultaneously. The number of these cases is relatively high, due to
the fact that many galaxies in our sample are located in or close to
cluster centers, where the galaxy density is high; {\it ii)} galaxies
with very low surface brightnesses, resulting in very messy profiles,
which could not be properly classified. Our final sample therefore
comprises 945 disk galaxies, among which we find 238 to be barred.

\subsection{The detectability of bars}\label{detect}
As shown by several studies \citep[e.g.,][]{kna99,esk00,mar07},
  the detectability of bars improves at longer wavelengths. The bar
  fraction measured for near-infrared observational studies are
  typically slightly higher than for studies
based on optical imaging. The main reason for this difference is
extinction caused by dust absorption, which is less severe at longer
wavelengths. Our sample of disk galaxies covers the redshift range
$z=0.4-0.8$, where the $I$-band observations correspond to rest-frame
$B$ to $V$. Hence, band-shifting effects should not be so strong, but
since the rest-frame range is rather blue, we should miss bars due to
dust absorption, and the provided bar fractions must be considered as
lower limits. In addition, enhanced star formation and dust
obscuration further impede bar detection at higher redshifts.

\begin{figure}
\centering
\includegraphics[width=9cm]{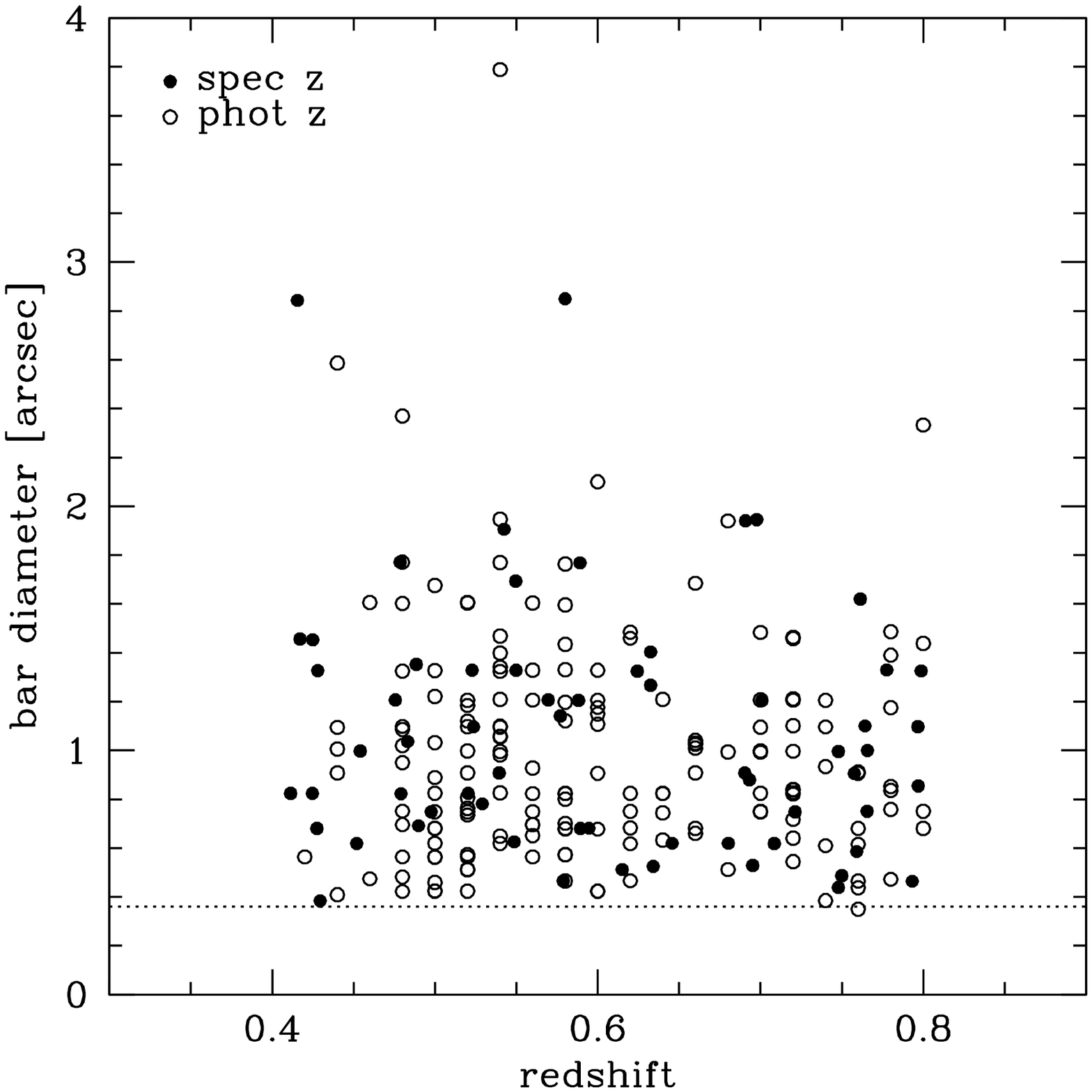}
\caption{Plot showing the bar diameter in arcsec versus redshift. The
  filled points indicate spectroscopically based redshifts, while open
  points are for photometric redshifts. The dotted horizontal line
  marks 4 PSFs (i.e. $0\farcs36$).}
\label{zblarc}
\end{figure}

\begin{figure}
\centering
\includegraphics[width=9cm]{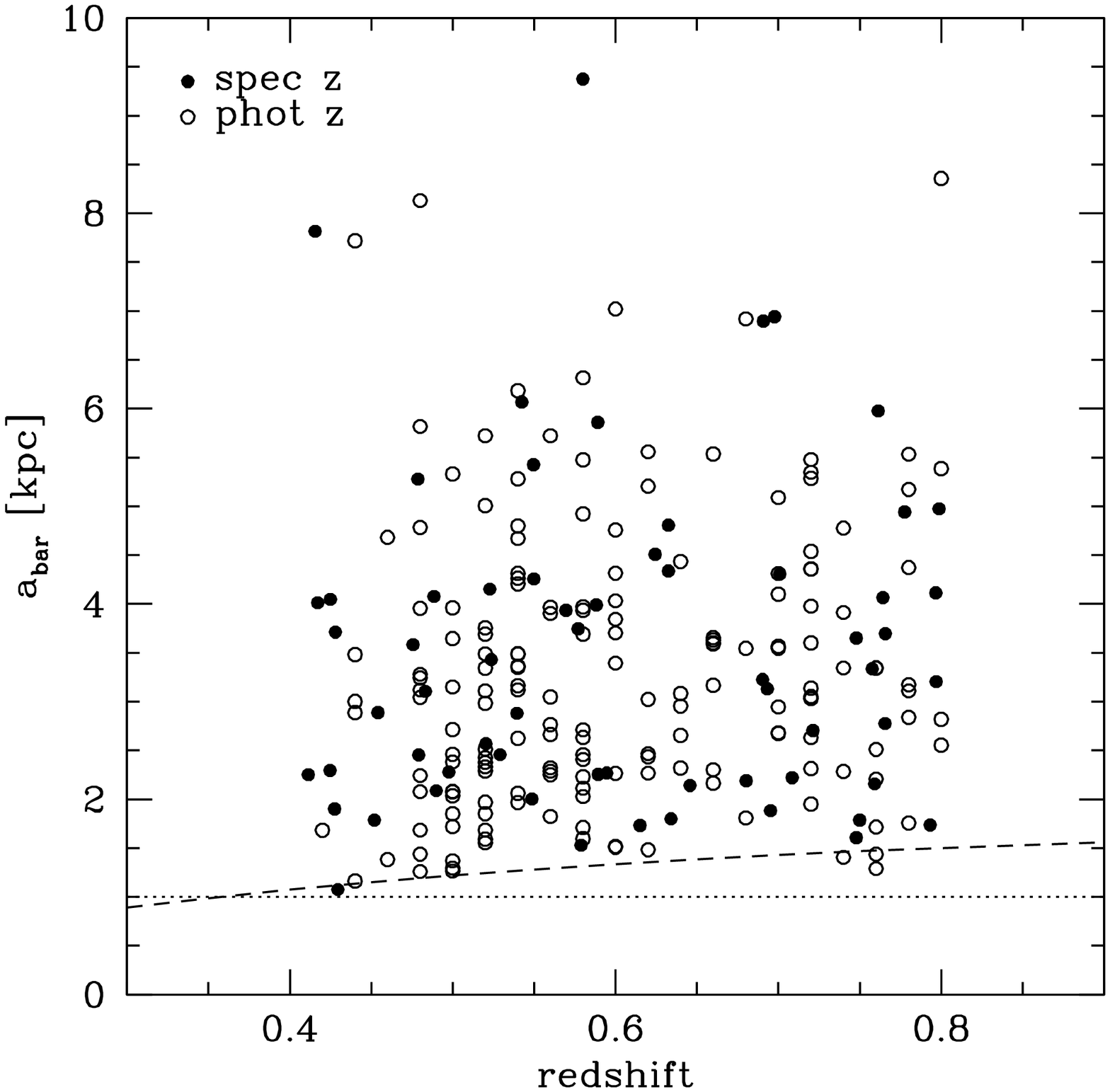}
\caption{The same as Fig. \ref{zblarc}, but this time showing the
  bar size in terms of bar semi-major axis $a_{bar}$. The dotted
  horizontal line shows the separation between large scale bars
  ($a_{bar}>1$ kpc) and nuclear bars ($a_{bar}\le1$ kpc). The dashed
  line indicates the absolute size of four pixels.}
\label{zbl}
\end{figure}

Another factor affecting the identification of bars in disk galaxies
is the resolution of the observations. We apply two requirements for
bar detection: (1) a bar can only be detected if its angular diameter
encompasses at least 4 PSFs; (2) the bar size (i.e., the bar radius,
$a_{bar}$) covers at least four pixels. Criterion (1) is based on the
fact that we need at least one PSF for the bulge, two PSFs for the bar
region, and one PSF for the disk beyond the bar. This resolution is
needed to detect reliably the quantitative bar signatures
described in Sect. \ref{method}. The PSF on our images
is $0\farcs09$, which corresponds to $\sim675$ pc at the highest
redshift ($z\sim1.0$) and 4 PSFs correspond to $\sim2.7$ kpc at that
redshift. In Fig. \ref{zblarc}, we plot redshift versus the bar
diameter in arcsec. The dotted line indicates 4 PSFs (i.e., $0\farcs36$).
The second requirement for bar detection is illustrated in Fig.
\ref{zbl}, where we plot $a_{bar}$ as a function of redshift. The
dotted line indicates the lower limit for large-scale bars at 1 kpc
\citep{lai02} and the dashed line corresponds to the absolute size of
four pixels. For increasing redshifts higher than $\sim0.4$, we begin
to lose the smallest bars. However, based on both bar-detection
criteria, we are complete for bars with $a_{bar}\ga2$ kpc.

Besides resolution, other factors, such as increased obscuration due
to both dust and star formation and surface-brightness dimming, can prevent
bar detection at higher redshifts. Small bars are particularly
affected by these factors. We point out that all results presented in
this study have been checked with regard to a possible bias with
respect to redshift (e.g., see Fig. \ref{effbl}).

\section{The optical bar fraction and its dependence on galaxy
  properties}\label{resul}
In the following, we discuss the properties of our disk galaxy sample
based on our bar classification. We show how the presence of a bar is 
related to other characteristics of the galaxies and whether there is
a difference between galaxies in clusters and the field. In general,
we use the sample of 945 moderately inclined disk galaxies for our
analysis, while when comparing cluster and field galaxies we restrict
the sample to objects with spectroscopic observations (136 galaxies in 
clusters and 105 galaxies in the field). This reduces the size of the
sample considerably, but ensures reliable membership assignment. In
particular, the contamination of the cluster sample by field galaxies
based on photometric redshifts impedes an accurate separation between
cluster and field galaxies. The error bars in the following plots 
  only include Poissonain errors. Specific factors affecting bar
  detection are discussed in Sect. \ref{detect}.

Among the 945 disk galaxies in our sample, we find 238 to be
  barred, and hence derive an optical bar fraction ($f_{bar}$) of $\sim25\%$.
  This is significantly lower than is typically found in optical
  studies of {\it local} galaxies \citep{esk00,mar07,ree07,bar08,agu09}, but
  in good agreement with studies at intermediate redshifts
  \citep{elm04,sht08}. This could indicate that the bar fraction is
  lower at higher redshifts, which would imply that we can detect
  nearly all bars in our sample. Possible other reasons for the
  difference compared to local studies have been given in Sect.
  \ref{detect}. We also point out that the
bar fraction for bulge-dominated galaxies is found to be significantly
lower than for disk-dominated systems in studies of local
  galaxies \citep{bar08,agu09}, a result
confirmed by our own study (Sect. \ref{bfmor}). Since our sample is
dominated by early-type disks ($>74\%$ are earlier than Sc), a
relatively low bar fraction might be expected. If we only consider strong
bars ($e>0.4$), we obtain a bar fraction of $\sim20\%$, which also
agrees well with earlier studies \citep{jog04}. For the
spectroscopically confirmed cluster sample, we obtain $f_{bar}=24\%$
(136 objects / 33 bars), and for the corresponding field sample, we derive
$f_{bar}=29\%$ (105/30). These values agree within the uncertainties
with the result for the complete sample and indicate that the frequency of
bars in clusters is almost identical to that in the field. This finding
indicates that bar formation, in general, is independent of environment,
which was also found by \citet{van02}. On the other hand, \citet{var04}
found twice as many bars in perturbed as in isolated galaxies. We
investigate these results further in Sect. \ref{discu}.

\subsection{The bar fraction as a function of morphological type}\label{bfmor}
\begin{figure*}
\centering
\subfigure
{
    \includegraphics[width=9cm]{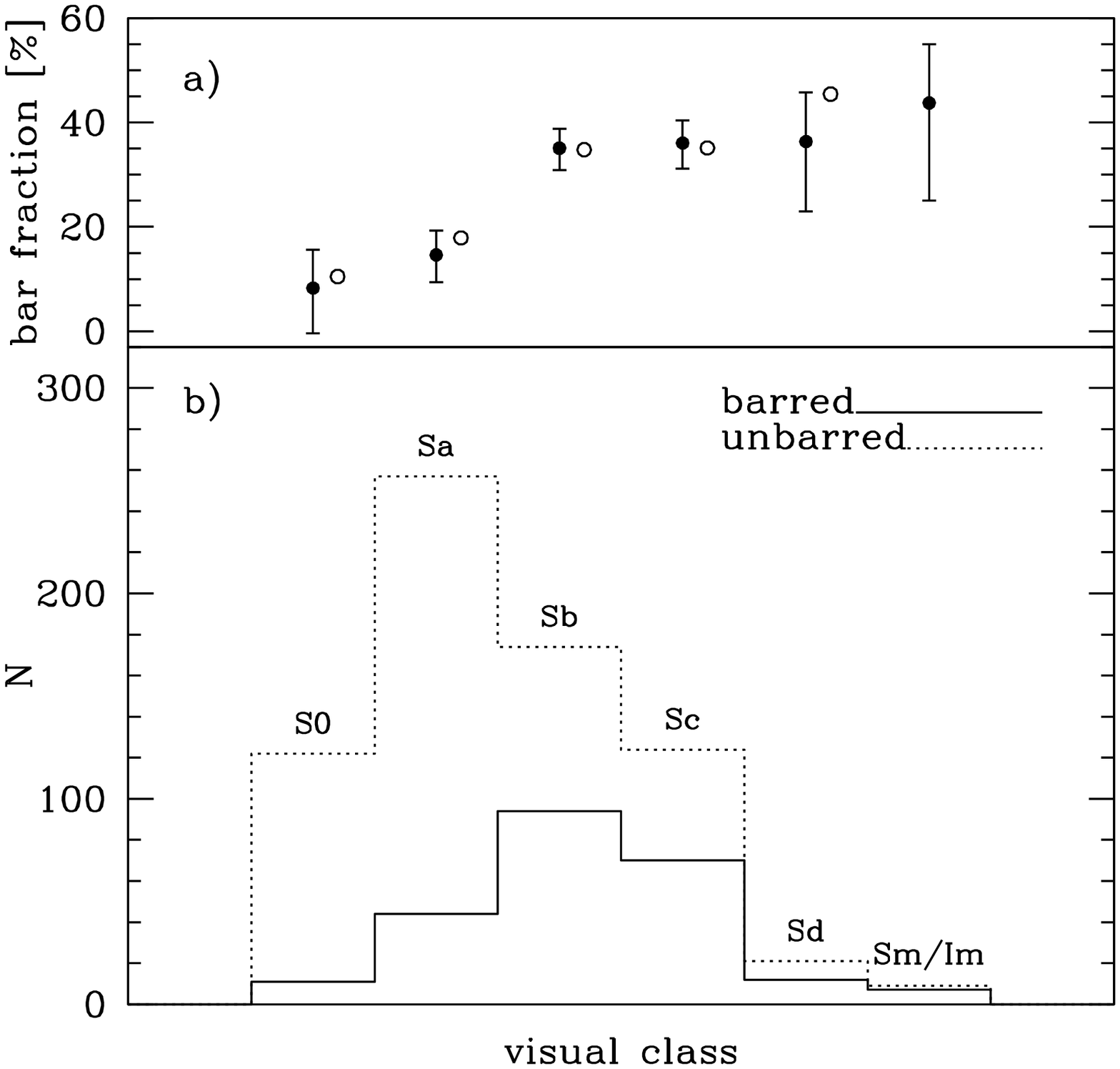}
}
\hspace{-0.5cm}
\subfigure
{
    \includegraphics[width=9cm]{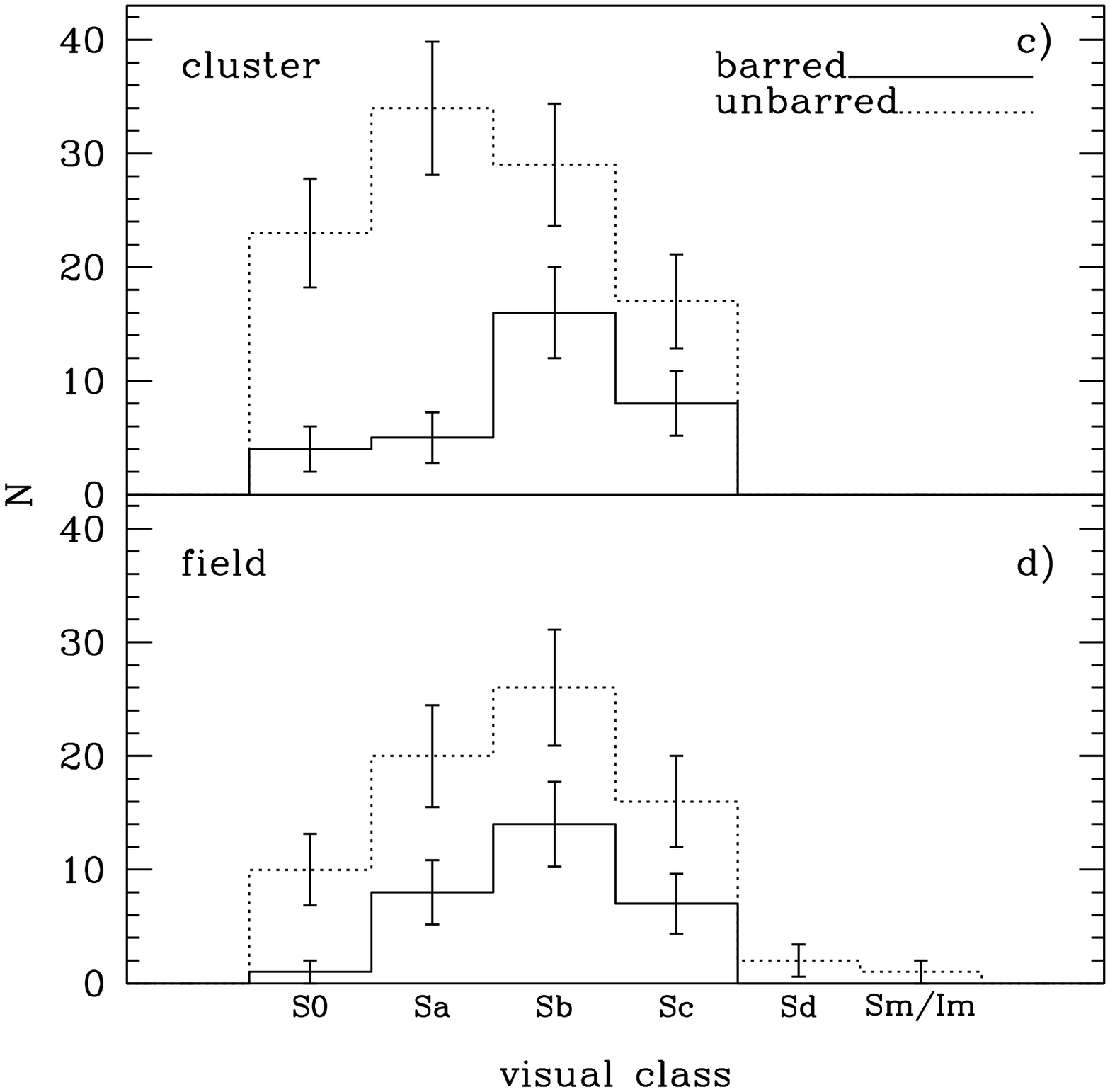}
}
\caption{The distribution of barred and unbarred disk galaxies as a
  function of Hubble type. {\it (a)} The bar fraction as a function of
  Hubble type. The filled circles show the relation for our final
  sample of 945 disk galaxies. The open circles show the relation for the
    complete sample with $M_V\leq-20$. {\it (b)} Histograms of barred
  (solid line) and unbarred (dotted line) disk galaxies. {\it (c)} The
  same as {\it (b)}, but for the spectroscopically based subsample of
  cluster members. {\it (d)} The same as {\it (b)}, but for the
  spectroscopically based subsample of field galaxies.}
\label{morph}
\end{figure*}

In Fig. \ref{morph}, we show the optical bar fraction as a function
of Hubble type. The results for the total sample (Figs. \ref{morph}a
and \ref{morph}b) indicate that the bar fraction increases towards
later Hubble types. Galaxies earlier than Sb have bar fractions below
$20\%$, while all later types exhibit higher bar fractions.
The monotonically increasing bar fraction along the sequence and the
significant difference between those of S0s ($f_{bar}=8\pm8\%$) and Scs
($f_{bar}=36\pm5\%$) indicate that the effect is significant. Our
results for Sd and Sm/Im types are less robust due to the smaller number
of galaxies with these morphologies. Similar analyzes
\citep{ode96,elm04} based on the Third Reference Catalog of Bright
Galaxies \citep{dev91}, however, obtained different results. Their bar
fraction was at its lowest for Sc galaxies and increased towards earlier and
later types. We cannot compare the results for S0 galaxies, since 
  these wre not included in the earlier investigations. On the
other hand, our result is consistent with two
studies based on the Sloan Digitized Sky Survey (SDSS), where,
however, no Hubble types were available: \citet{bar08} showed that the
bar fraction is significantly higher for disk-dominated galaxies than
for galaxies hosting prominent bulges. Although the sample in that
study was dominated by late-type disks, the connection found between
the prominence of the bulge and the bar fraction can also be inferred
from our result. In their analysis of bars in local disk galaxies,
\citet{agu09} found that S0 galaxies have a significantly lower bar
fraction (by $\sim23\%$) than galaxies of later types, which
agrees well with our result. A somewhat different result was
  reported by \cite{sht08}, who found a slight preference for bars in
  bulge--dominated systems at high redshifts. We emphasize that the presence of a
large bulge should not impede bar detection, since we are only
interested in large--scale bars and exclude strongly inclined objects,
for which a massive bulge could make bar detection difficult.

Figures \ref{morph}c and \ref{morph}d show the same relations for our
spectroscopic sample. The distributions indicate that early-type disks
are more prominent in clusters, as expected, whereas the bar fractions
in the field and in clusters are quite similar. On the other hand, we find
differences in the bar fractions for individual Hubble types in
clusters and the field, for instance for Sa galaxies (cluster:
  $13\%$, field: $29\%$). However, due to the small number of objects
involved (20--40 galaxies), the significance of this result cannot be
assessed reliably.

\subsection{The bar fraction as function of magnitude and color}
\begin{figure*}
\centering
\subfigure
{
    \includegraphics[width=9cm]{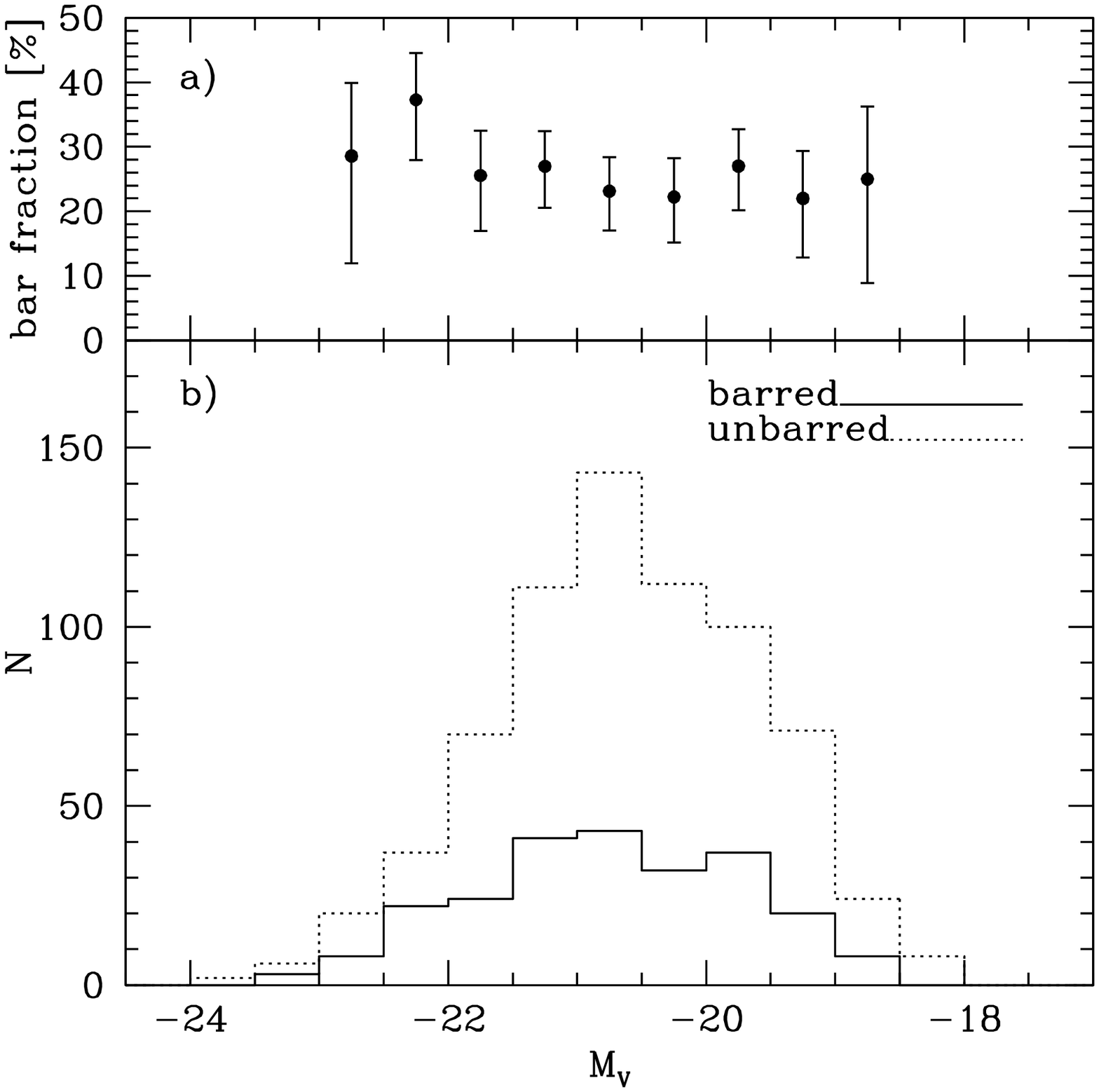}
}
\hspace{-0.5cm}
\subfigure
{
    \includegraphics[width=9cm]{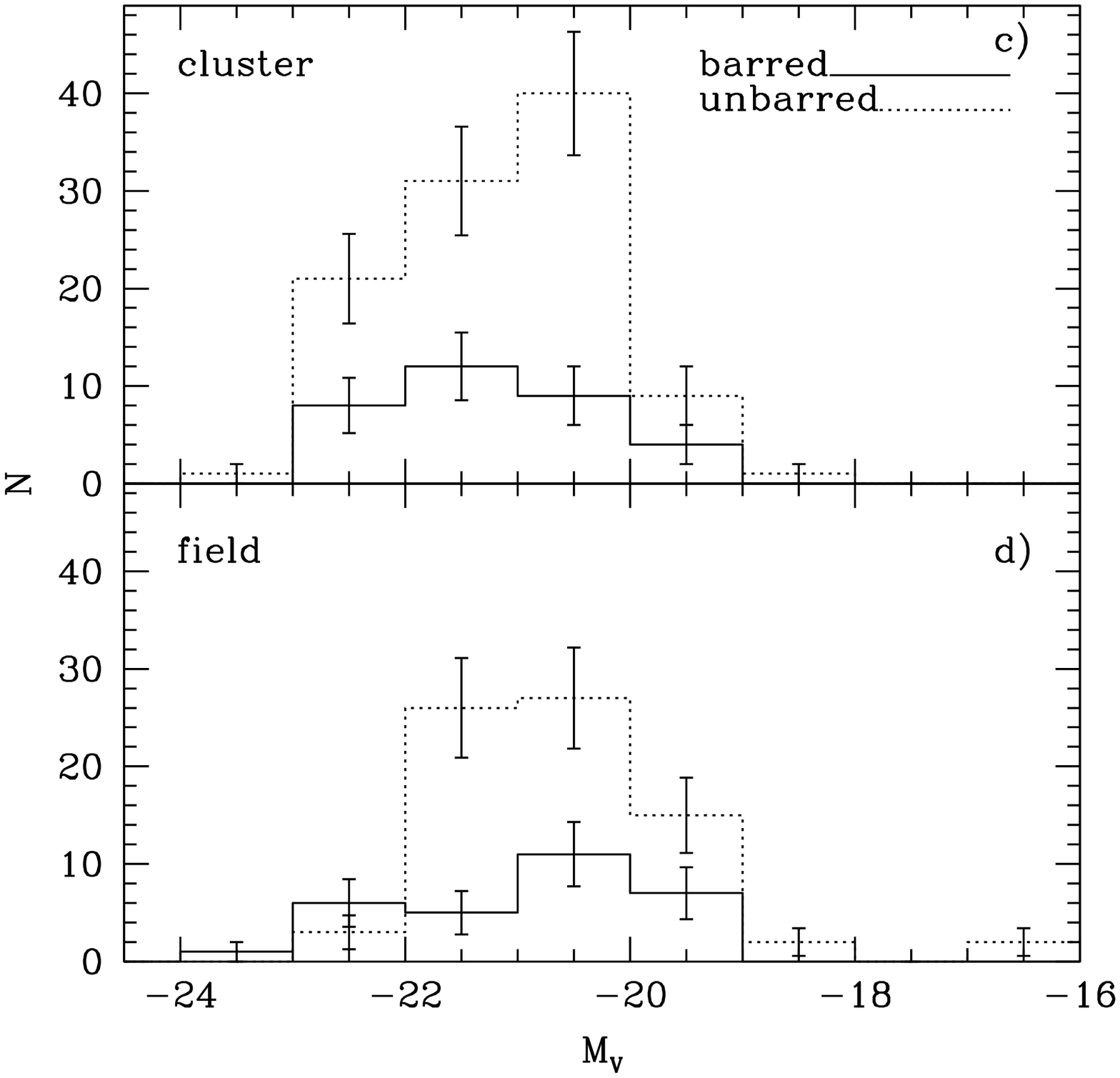}
}
\caption{The same as Fig. \ref{morph}, but for the absolute $V$
  magnitude. In {\it (a)} only bins with more than 25 objects are shown.}
\label{mag}
\end{figure*}

In Fig. \ref{mag}, we show how the optical bar fraction depends on
the rest-frame total $V$-band magnitude of the galaxy, for both the full
sample and the spectroscopic subsample. For cluster members, the
rest-frame magnitudes are calculated based on the cluster redshift (as
opposed to the galaxy redshift).
The bar fraction remains relatively constant with galaxy
magnitude. The slight decrease towards lower magnitudes lies within
the uncertainties and is therefore insignificant. Early-type disks
are generally more luminous than late-type disk and have a lower bar
fraction (Fig. \ref{morph}). We may therefore naively expect that
the bar fraction should increase toward lower magnitude. This trend
could be erased, however, if the brighter galaxies in each
morphological class had a higher bar fraction, which has been
  found in a study of barred disks in the Abell 901/902 supercluster
  environment \citep{mar09}. To
investigate this possibility, we split the morphological subsamples at
$M_V=-21.0$ mag, which is the mean $V$-band magnitude of the complete
sample ($M_V\leq-20.0$), and determine the bar fractions for the brighter and fainter
parts of these subsamples. The result is shown in Table \ref{t_morph}.
\begin{table}
\caption{Bar fractions per Hubble type for bright and faint subsamples}
\label{t_morph}
\centering
\begin{tabular}{c c c}
\hline
Hubble & \multicolumn{2}{c}{Bar fraction (Number of objects)} \\
type & $<-21.0$ mag & $\ge-21.0$ mag \\
\hline
 S0 & $14\%$ (44) & $6\%$ (89) \\
 Sa & $21\%$ (121) & $11\%$ (180) \\
 Sb & $40\%$ (117) & $31\%$ (151) \\
 Sc & $32\%$ (59) & $38\%$ (135) \\
\hline
\\
\end{tabular}
\begin{minipage}{\columnwidth}
Notes: The table shows results for 896 objects in the range
$0.4<z<0.8$. (The 52 Sd/Sm/Im galaxies are not considered in this
Table.) The samples are split at $M_V=-21.0$ mag, which is the mean
$V$-band magnitude of the total sample.
\end{minipage}
\end{table}
For Hubble types S0--Sb, the bright subsample does indeed have a higher bar
fraction than the faint subsample. For Sc disks, the
trend is, however,  reversed. This explains the almost constant bar fraction as a 
function of magnitude. The result indicates that most bars in
early-type disks have a relatively high surface brightness and
therefore contribute significantly to the high luminosity.

Figures \ref{mag}c and \ref{mag}d show the distributions for
the cluster and field subsamples. The cluster sample is slightly
brighter than the field sample, but in terms of bar fraction they are
almost identical.

\begin{figure*}
\centering
\subfigure
{
    \includegraphics[width=9cm]{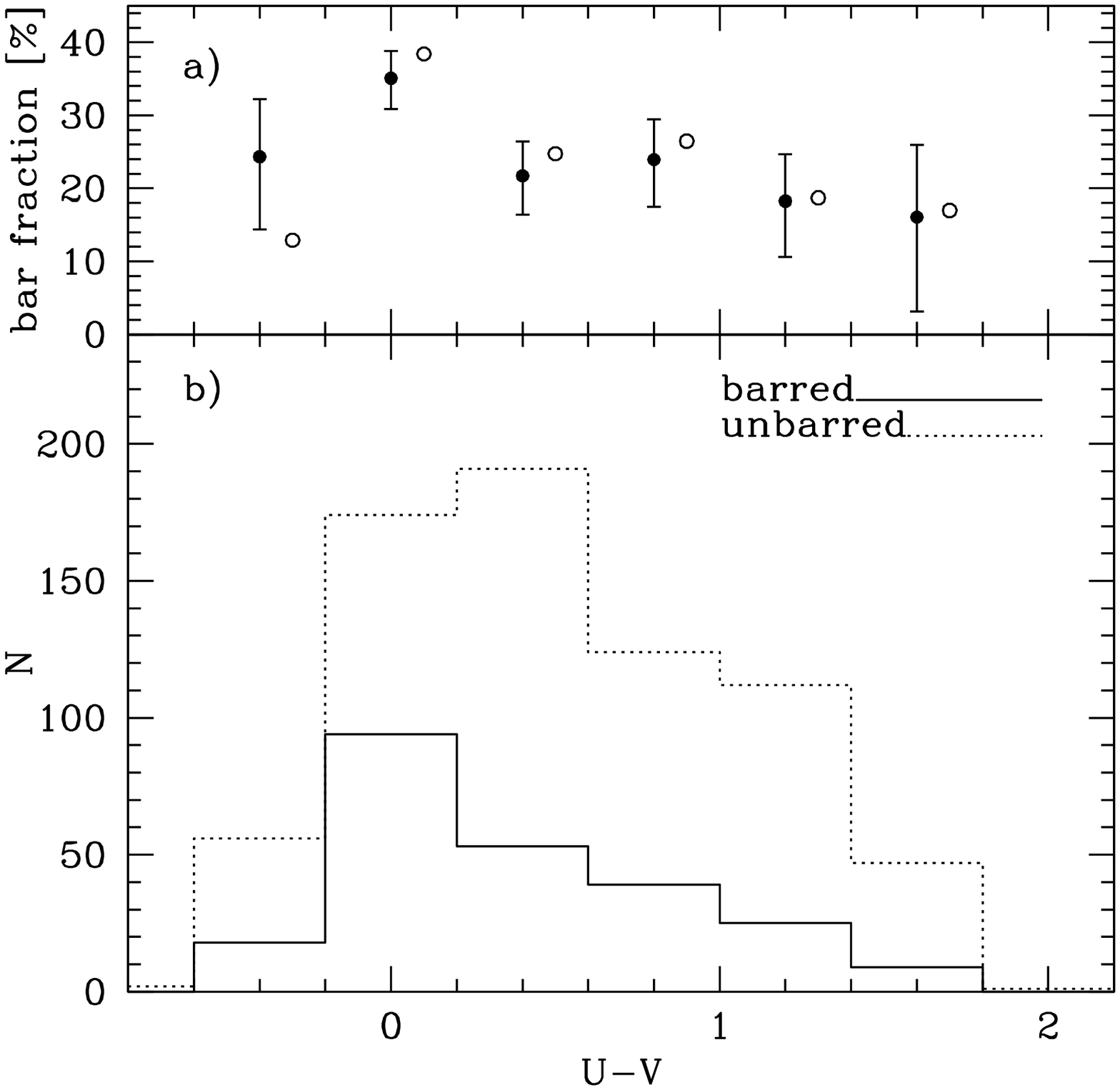}
}
\hspace{-0.5cm}
\subfigure
{
    \includegraphics[width=9cm]{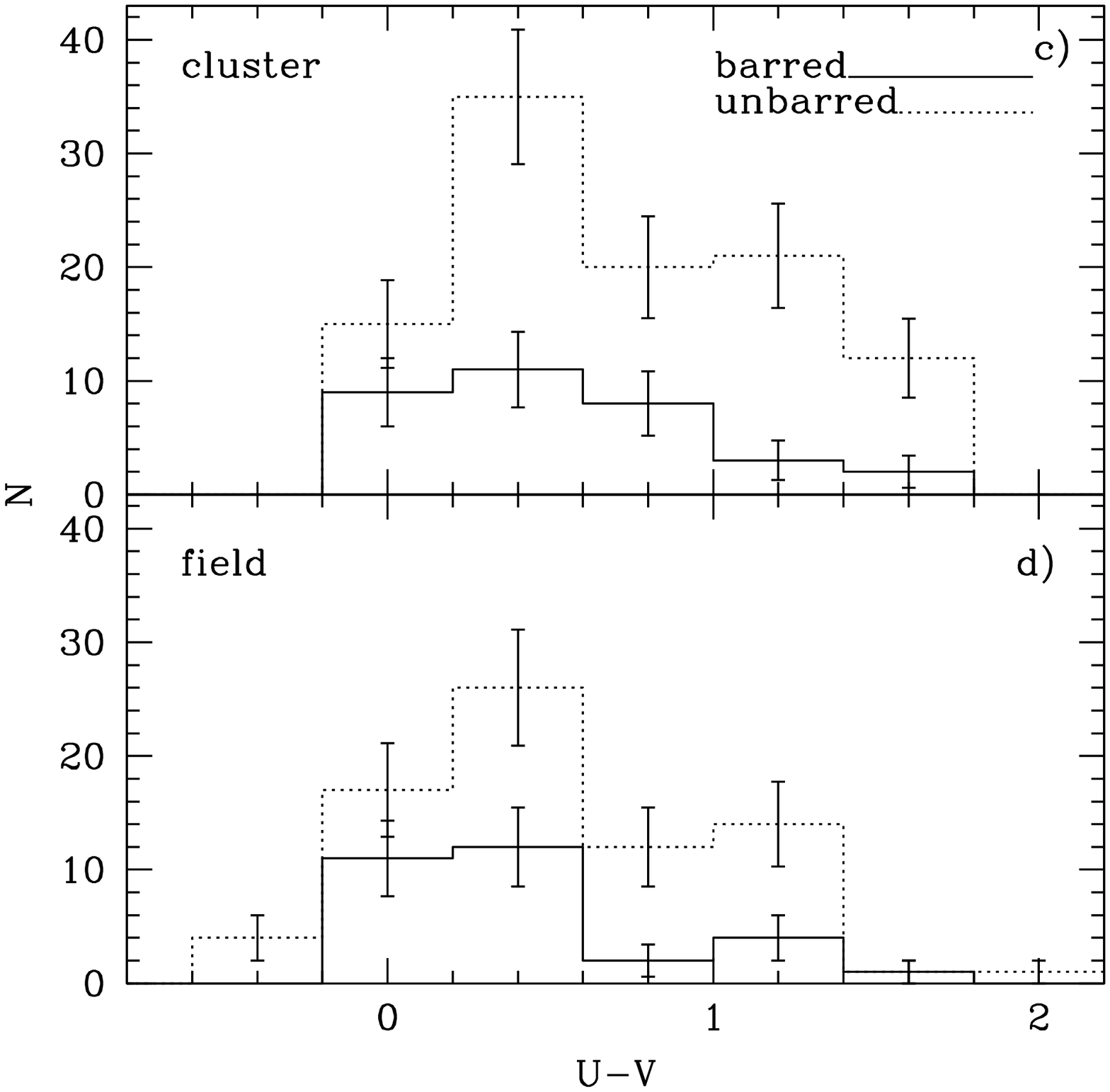}
}
\caption{The same as Fig. \ref{mag}, but for rest-frame $U-V$ color.}
\label{color}
\end{figure*}

The corresponding relations with respect to rest-frame $U-V$ color are
shown in Fig. \ref{color}. The bar fraction declines toward redder
colors (Figs. \ref{color}a and \ref{color}b), as expected from the
relations found in terms of morphology. This finding emphasizes that
late-type disks are more likely to host bars than early-type
disks. The same trend is also found for the cluster and field
subsamples (Figs. \ref{color}c and \ref{color}d), although the
cluster sample is on average significantly redder than the field
sample.

\subsection{The bar fraction as a function of effective radius}\label{seff}
\begin{figure*}
\centering
\subfigure
{
    \includegraphics[width=9cm]{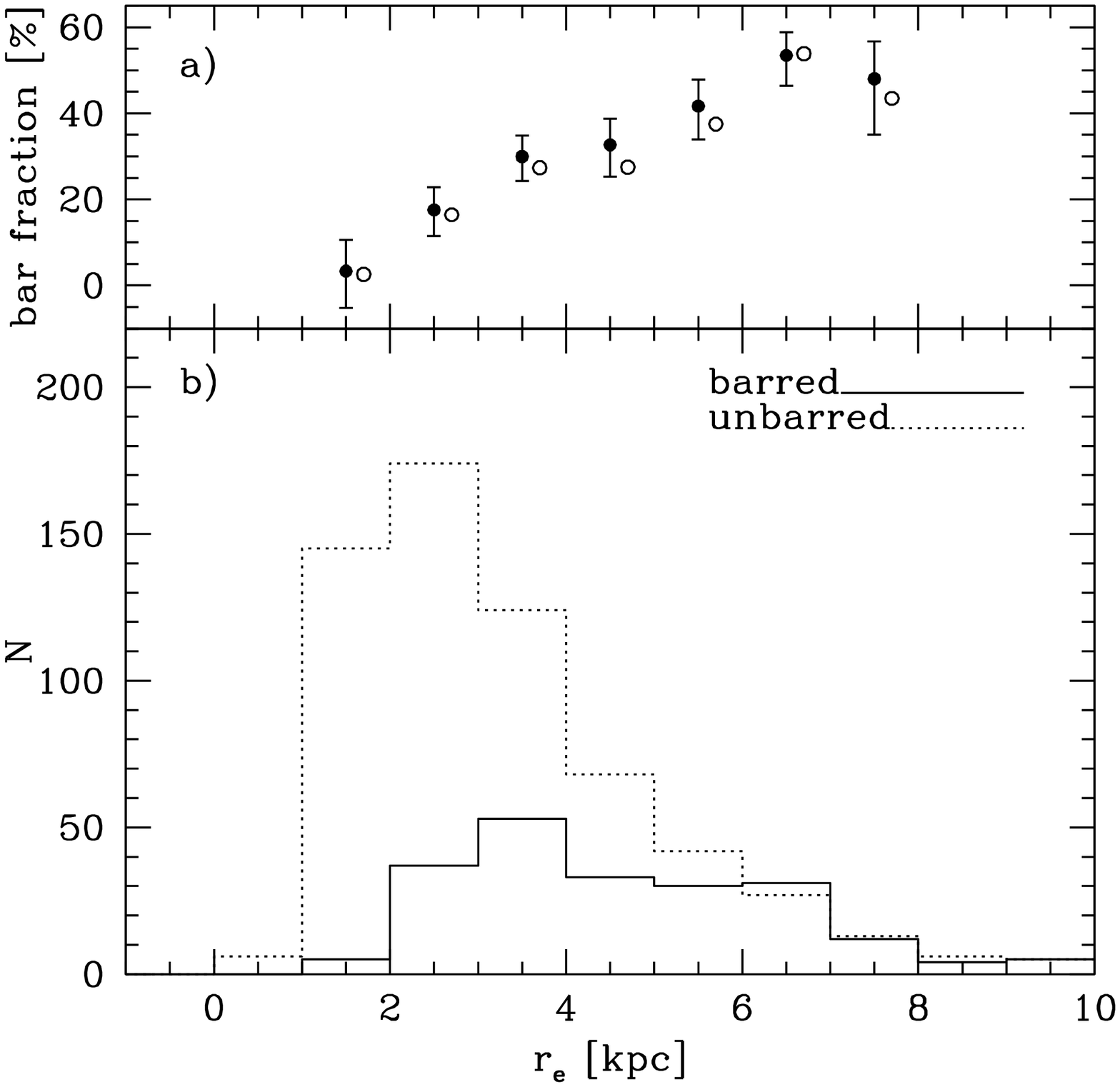}
}
\hspace{-0.5cm}
\subfigure
{
    \includegraphics[width=9cm]{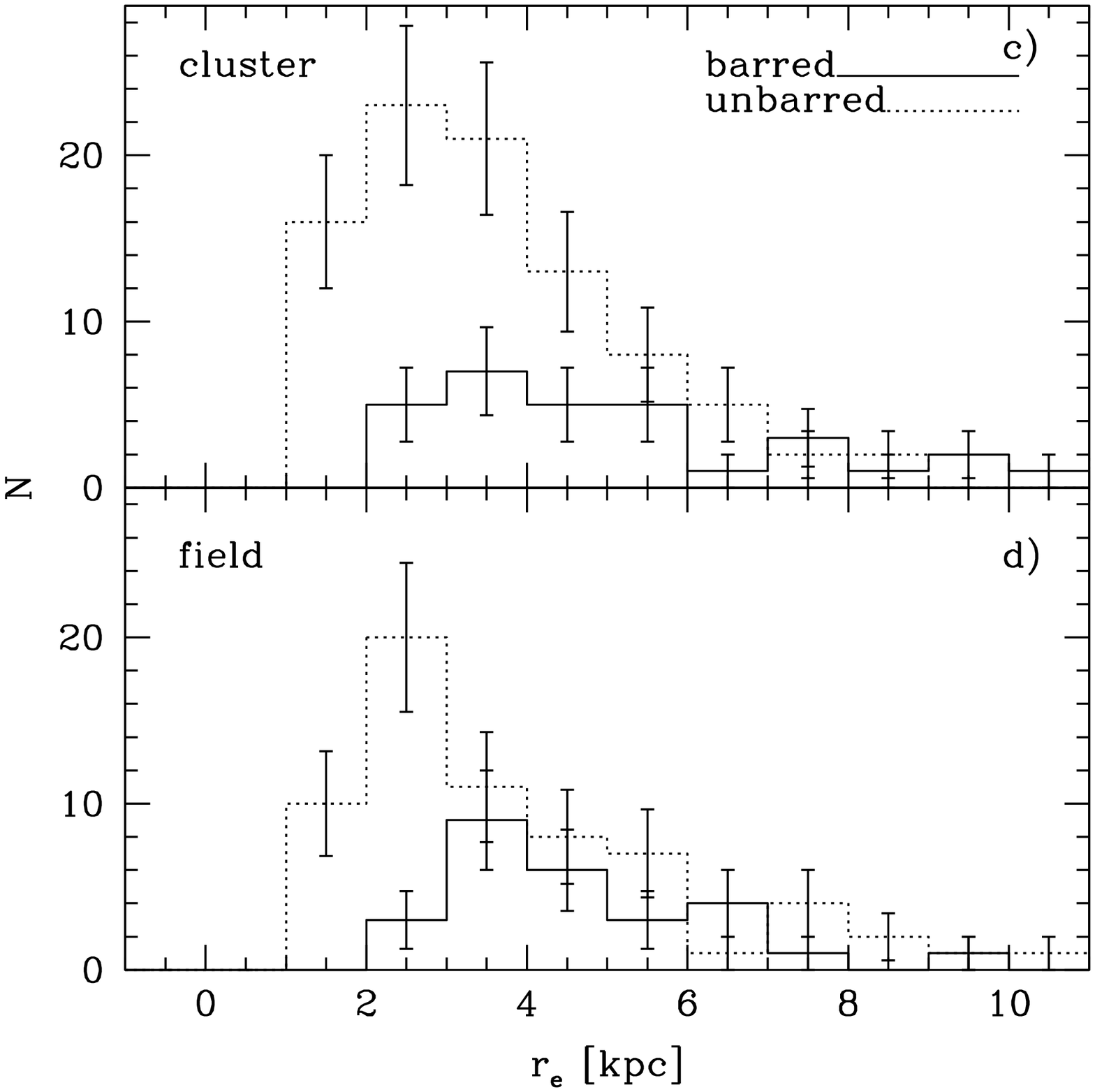}
}
\caption{The same as Fig. \ref{mag}, but for the effective radius
  determined from a S\'ersic fit. Galaxies in the clusters
  cl1227.9-1138 and cl1227.9-1138a are excluded, since S\'ersic fits
  are unavailable.}
\label{eff}
\end{figure*}

\begin{figure}
\centering
\includegraphics[width=9cm]{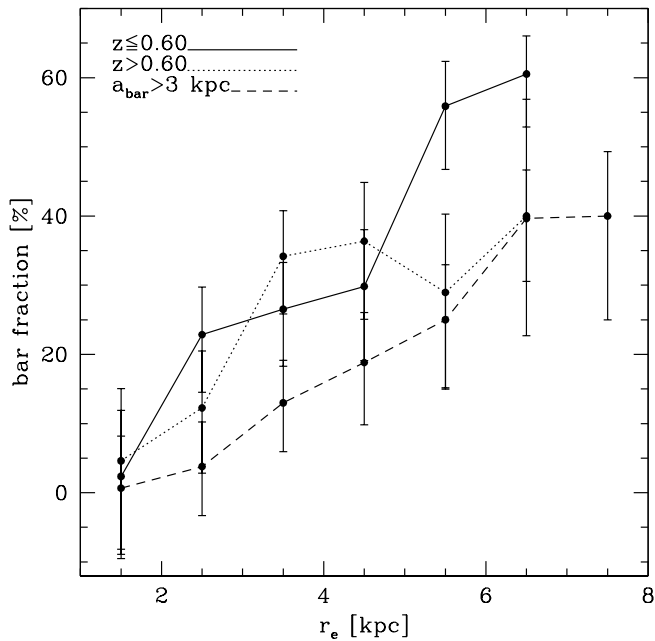}
\caption{The same as Fig. \ref{eff}a, but for three different
  subsamples: the solid line is for galaxies at $z\le0.60$ (the median 
  redshift of the sample), the dotted line is for galaxies at
  $z>0.60$, and the dashed line shows the relation assuming that only
  bars with $a_{bar}>3$ kpc can be detected.}
\label{effbl}
\end{figure}

The effective radius ($r_e$) determined by applying a S\'ersic fit to the
entire galaxy is a measure of the concentration of the galaxy light and 
a partial measure of the prominence of the bulge. A relation
  between Hubble type and $r_e$ is therefore expected and also found
  in our sample. We computed the mean effective radius for each
  Hubble type and obtain the following result: S0: $\langle r_e
  \rangle=1.81$ kpc; Sa: $\langle r_e \rangle=2.46$ kpc; Sb: $\langle
  r_e \rangle=3.54$ kpc; Sc: $\langle r_e \rangle=4.27$ kpc; and Sd:
  $\langle r_e \rangle=3.61$ kpc. However, since $r_e$ only indicates
  the distribution of the light in the galaxies, it is more related to
  the concentration of the light than the Hubble type. Figure \ref{eff}
shows the bar fraction as a function of $r_e$. There is a continuous
rise in the bar fraction with increasing $r_e$ (Fig.
\ref{eff}a). This again seems to be just another representation of the
effect already indicated in the relations based on morphology and
color. Galaxies with larger central light (or mass) concentrations
have less bars than disk-dominated galaxies with small central
objects. The result also agrees with the findings
of \citet{bar08} where a similar plot is shown and with the results
of \cite{her08}, which are based on the light concentration index.
However, the apparent lack of the smallest
bars at higher redshifts might contribute to this result. This could
be the case if galaxies with small effective radii tend to have
shorter bars. We therefore reproduce Fig. \ref{eff}a for three
different subsamples. The result is shown in Fig. \ref{effbl}. The
solid line is for galaxies at $z\le0.60$ (the median redshift of the
sample), the dotted line is for galaxies at $z>0.60$, and the dashed
line shows the relation assuming that only bars with $a_{bar}>3$ kpc
can be detected (i.e., all bars with $a_{bar}<3$ kpc are considered to be
unbarred in this case). All subsamples clearly show that the bar
fraction increases towards galaxies with larger effective
radii. We can therefore conclude that the result is real and not
caused by the lower  bar detection rate at higher redshift.

Finally, the same result is again found in the two subsamples in
Figs. \ref{eff}c and \ref{eff}d. This finding supports the assumption
that the presence of a bar is related to the disk structure and the
magnitude of the central mass concentration.

\section{The properties of the bars}\label{bprop}
The two main properties of bars typically measured and
analyzed in bar studies are the bar size and strength. Different
  methods are adopted to determine these quantities. In particular,
the bar strength can be measured using the gravitational torque
exerted by the bar \citep{blo02,but05}, the maximum ellipticity of the 
bar \citep{mar95,jog99,kna00,lai02,mar07,bar08}, Fourier decomposition 
\citep{elm96}, and visual estimates of strength \citep{esk02}. As
described in Sect. \ref{method}, our measurements are based on ellipse
fitting. Since this fitting provide the basis of the classification, the 
determination of the maximum ellipticity ($e_{bar}$, interpreted as
the bar strength) and the radius ($a_{bar}$, interpreted as the bar
size), at which this maximum occurs, is straightforward. However, we
emphasize that the maximum bar ellipticity is only a partial
measure of the bar strength, because it provides no indication
of the mass of the bar. However, \citet{she04} demonstrated
that the maximum ellipticity correlates strongly with the relative
amplitude of the bisymmetric Fourier component of the mass density
averaged over a certain inner radial range, where the bar
dominates. In addition, the gravitational torque, on average,
correlates with the maximum ellipticity of the bar \citep{lau02}. We
can therefore regard the bar ellipticity to some extent as a measure
of the bar strength.
\begin{figure}
\centering
\includegraphics[width=9cm]{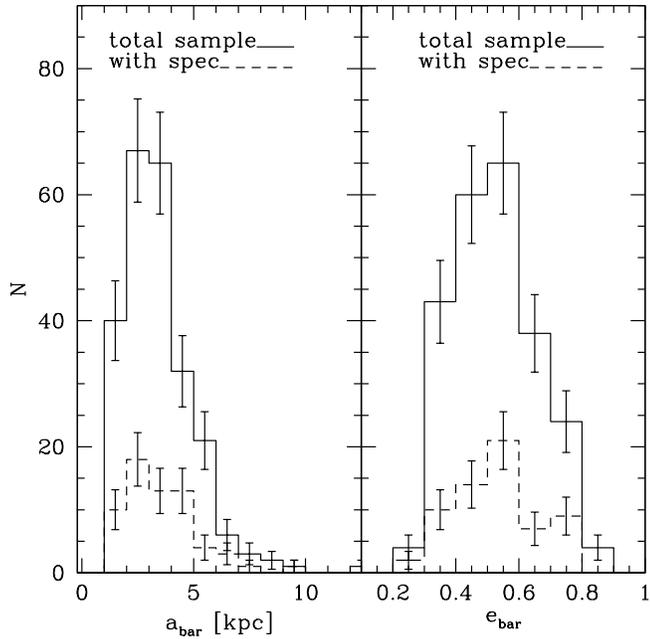}
\caption{The bar size (left) and bar ellipticity (right) distributions 
  for the total sample (solid lines) and the spectroscopic sample
  (dashed line).}
\label{blbs}
\end{figure}

\begin{figure}
\centering
\includegraphics[width=9cm]{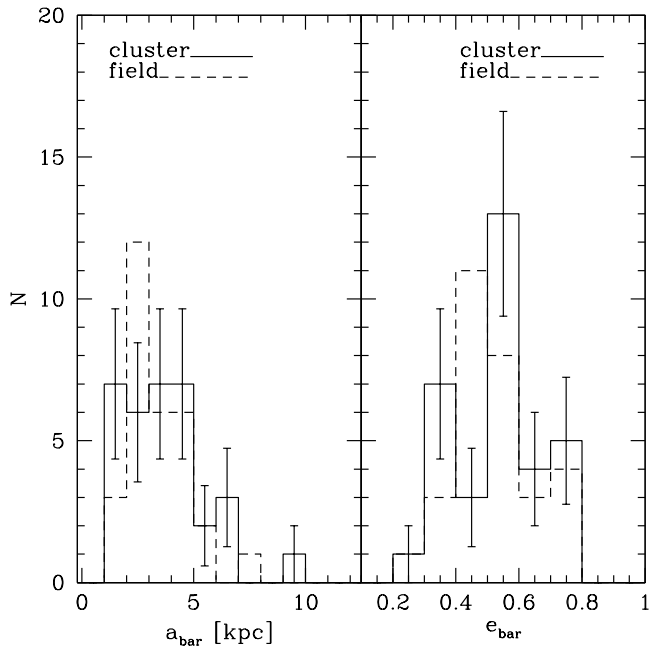}
\caption{The same as Fig. \ref{blbs}, but only showing the
  spectroscopically defined cluster (solid lines) and field (dashed
  lines) subsamples. For readability, error bars are only shown for
  the cluster sample.}
\label{blbsss}
\end{figure}
In Fig. \ref{blbs}, we show the bar size and bar ellipticity
distributions. A significant majority ($86\%$) of bars have sizes $\le5$ kpc,
in agreement with un-deprojected results in earlier studies
\citep{jog04,mar07,bar08}. Similarly, the distribution of $e_{bar}$
is consistent with these earlier studies. Figure \ref{blbsss} shows
the corresponding distributions for the cluster and field
subsamples. The cluster sample exhibits no prominent peak in
the bar size distribution (left panel), but is slightly skewed towards
larger sizes. In contrast, the field sample shows a similar
distribution to that of
the total sample. Hence, bars in clusters tend to be longer than bars
in field galaxies.

\begin{figure}
\centering
\includegraphics[width=9cm]{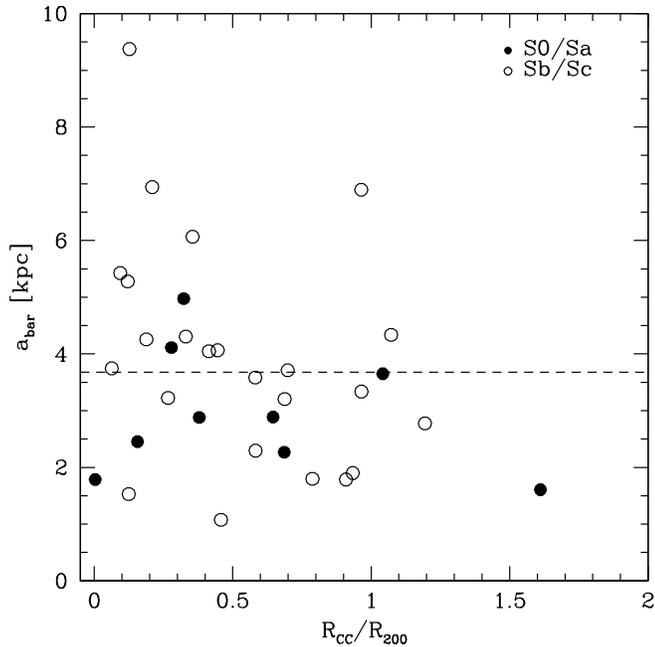}
\caption{The bar size as a function of normalized clustercentric
  distance for the spectroscopic cluster subsample. Filled points are
  for S0/Sa types and open point for Sb/Sc types. There are no other
  Hubble types in the sample. The dashed line indicates the mean bar
  size for this sample of 3.68 kpc.}
\label{blcc200}
\end{figure}

The differences that we observe between the bar size distributions of
the field and cluster samples (Fig. \ref{blbsss}) could indicate
that the cluster environment has an impact on the bar sizes. To
investigate this possibility, we measured the projected distances
($R_{\rm CC}$) of the cluster galaxies from the corresponding cluster
center, assuming that all galaxies are at the cluster redshift. This
value was then normalized by $R_{\rm 200}$, the radius within which
the average mass density is equal to 200 times the critical
density. We use the definition given in \citet[][equation
8]{fin05}. In Fig. \ref{blcc200}, we plot the bar size versus the
normalized clustercentric distance. Larger bars tend to be located
close to the cluster center. The majority of bars larger than the
average size of 3.68 kpc (indicated by the dashed line) are located
within $R_{CC}/R_{200}<0.5$. A KS-test indicates that there is a $\sim1\%$
probability that the $a_{bar}>3.68$ kpc sample stems from the same
distribution as the complementary sample of small bars in terms of
$R_{CC}/R_{200}$. We must remember that we are using projected
distances, although this is unlikely to explain the trend entirely.
It is also interesting that
most of the $a_{bar}>3.68$ kpc bars are found in disks of intermediate 
Hubble type. However, we highlight that the number of objects
(33) in the plot is small and the significance of this
finding is difficult to assess. We will nevertheless discuss
possible origins of this effect in Sect. \ref{discu}.

\begin{figure}
\centering
\includegraphics[width=9cm]{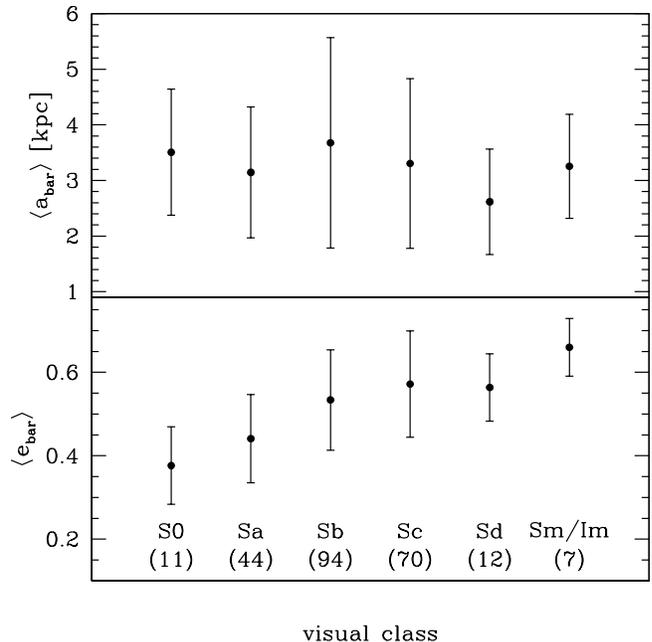}
\caption{The average bar size (top) and average bar ellipticity
  (bottom) as a function of Hubble type. The number in brackets
    gives the number of objects in that bin. The error bars indicate
    standard deviations of the mean.}
\label{barmor}
\end{figure}

In Fig. \ref{barmor}, we show how the bar properties are related to
Hubble type. The two panels show the mean bar size and bar ellipticity
for each Hubble type. The mean bar sizes are in the range 2.5--3.5
kpc. For early-type disks (S0--Sb), this is consistent with the results
of \cite{erw05}, whereas the mean bar sizes in late-type disks are
larger than those measured by \cite{erw05}. However, the number of objects in
the Sd and Sm/Im classes is rather small in our sample (see Fig.
\ref{morph}). On the other hand, there is an obvious relation between
morphological type and mean bar ellipticity, which increases by more than
0.2 from early to late types. However, this result is at least in part
due to the measurement including the bulge, which can
lead to an underestimation of the bar ellipticity \citep{gad08}. The
measured bar ellipticity stems from the outermost isophote attributed
to the bar. The bulge light contributes to the central part of this
isophote. A large and luminous bulge will therefore widen this isophote
at the center and, hence, lower the bar ellipticity. With the data
available, it is impossible to estimate the impact that this effect
could have
on the relation in Fig. \ref{barmor} (lower panel). An accurate
bulge-bar-disk decomposition would be required to investigate the
possibility that late-type disks indeed have stronger bars, on
average. This is, however, beyond the scope of this analysis.

\section{The distribution of barred galaxies within the clusters}\label{cldis}
\begin{figure}
\centering
\includegraphics[width=9cm]{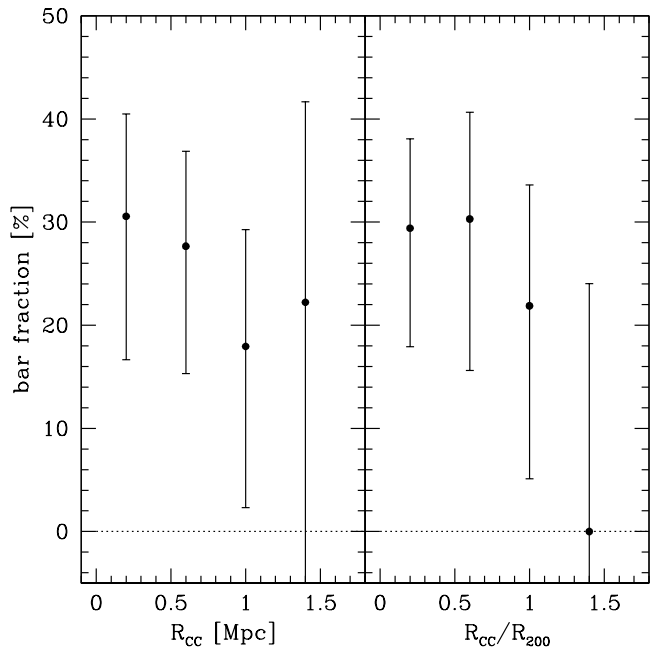}
\caption{The bar fraction as a function of the
  clustercentric distance (left) and the normalized clustercentric
  distance (right) for the spectroscopic subsample.}
\label{cdist}
\end{figure}

In two studies of local galaxy clusters, evidence has been found that
barred galaxies are preferentially located at the cluster core. An
analysis of the clustercentric distances of barred galaxies in the
Coma cluster \citep{tho81} showed that a significantly larger fraction 
of barred galaxies are located at the cluster core than larger clustercentric
distances (the adopted core radius for Coma was $\sim784$
kpc). In a similar study of the Virgo cluster, \citet{and96} found
that the barred disk galaxies are more centrally concentrated than the 
unbarred disks. For S0 galaxies, the same study found that the distributions of barred and
unbarred objects are the same. In Fig. \ref{cdist}, we show the
bar fractions as a function of the clustercentric distances (absolute and
normalized). We find the highest
bar fraction in the central bin. For the $R_{CC}$ distribution,
the bar fraction declines from $31\%$ in the central bin to $18\%$ at
$\sim1$ Mpc. (The corresponding values for the complete sample
  are $30\%$ and $15\%$, respectively.) For the $R_{CC}/R_{200}$ distribution, the corresponding
values are $29\%$ in the central bin and $22\%$ at
$R_{200}$. (The corresponding values for the complete sample
  are $30\%$ and $21\%$, respectively.) We emphasize again
that the sample used is rather small, but we can safely
say that barred galaxies do not avoid the cluster
center. With the tendency for galaxy with large bars to be
  located close to cluster cores (Fig. \ref{blcc200}), these
findings indicate that regions of high galaxy density are favorable
locations for bars.

\section{Discussion}\label{discu}
We have found the optical bar fraction averaged over our entire sample
covering the redshift range $z=0.4-0.8$ to be $\sim25\%$. The median
redshift of the sample is 0.60. In Sect. \ref{detect}, we discussed
how at these redshifts certain factors, such as reduced resolution,
band shifting, enhanced obscuration by dust, could lead to a reduced bar
detection rate. This could explain why the optical bar fraction of this
sample is considerably lower than measured in local samples of disk
galaxies \citep{mar07,bar08,agu09}, but is in good agreement with
studies at intermediate redshifts \citep{elm04,jog04,sht08}. We note
that all of these studies were based on samples primarily composed of
galaxies in low density environments. A key question is whether the
lower bar fraction at intermediate redshifts is primarily a result of
the enhanced difficulty in identifying bars, or whether the number of
barred disk galaxies was really lower at these redshifts. Several
studies \citep{abr99,van00} claimed that the bar fraction was
significantly lower at $z\ga0.5$. This was confirmed by
\cite{sht08}. Only for a small subsample of large
  and massive galaxies was a constant bar fraction found
  \citep{sht03,sht08}. In other studies, a fairly constant bar
fraction out to $z\sim1$ was found for strong bars \citep{jog04}
or all bars \citep{elm04,zhe05}. To determine the evolution
in the bar fraction with redshift in our sample, we divided the sample
into three redshift bins and measured the bar fraction and the mean bar
sizes and ellipticities in each. The bin sizes were defined in order to
obtain comparable numbers of objects in each bin. We only considered
galaxies with spectroscopic redshifts. The corresponding results are
shown in Table \ref{t_evol}.
\begin{table}
\caption{Bar fractions and properties as a function of $z$}
\label{t_evol}
\centering
\begin{tabular}{c c c c}
\hline
Spec-z & $0.4<z\le0.55$ & $0.55<z\le0.7$ & $0.7<z\le0.8$ \\
\hline
\# of objects & 87 & 78 & 76 \\
bar fraction & $30\%$ & $26\%$ & $22\%$ \\
$\left< a_{bar}\right>$ [kpc] & 3.38 & 3.83 & 3.37 \\
$\left< e_{bar}\right>$ [kpc] & 0.54 & 0.53 & 0.52 \\
\hline
\\
\end{tabular}
\begin{minipage}{\columnwidth}
Notes: The bar fraction and average bar sizes and ellipticities in
  three spectroscopically based redshift bins.
\end{minipage}
\end{table}
The bar fraction was found to decline modestly from $30\%$ at $0.4<z\le0.55$
to $22\%$ at $0.7<z\le0.8$. The average bar sizes in the lowest
  and highest redshift bins are roughly the same, and the average bar
  strength does not change within our redshift range. This indicates
  that we have not missed significant numbers of short or weak bars at
  higher redshifts. These results suggest that either the bar
fraction decreases with increasing redshift, which would
imply that we can nearly detect all bars in our sample, or that the
decline is caused by the growing difficulty in identifying bars at
higher redshifts. Reasons for the latter were given in Sect.
\ref{detect}. A detailed analysis of this issue was beyond the scope of 
this study.

The other results presented in the previous sections can be divided
into two categories: (1) general relations between bars and the
properties of their host galaxies, independent of whether the galaxies
are cluster members or in the field; (2) relations regarding the
specific locations of barred galaxies within the clusters.

\subsection{Relating bar fraction and characteristics to host galaxy
  properties}
\begin{figure}
\centering
\includegraphics[width=9cm]{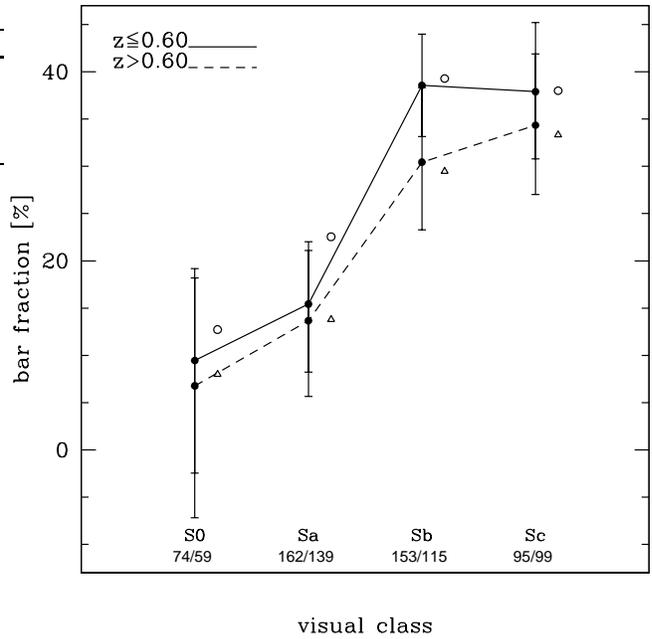}
\caption{The bar fraction as a function of Hubble type, separated into 
  a subsample at $z\le0.60$ (solid line and open circles for the
  complete sample) and a subsample at $z>0.60$ (dashed line and
    open triangles for the complete sample). The numbers at the
  bottom of the plot indicate: number of galaxies at $z\le0.60$/number
  of galaxies at $z>0.60$. We omit Hubble types Sd and later, since
  the number of objects in these  bins is too low. The error bars only
  include Poissonain errors.}
\label{morph_z}
\end{figure}

We have found additional evidence that the bar fraction is related to the
morphological structure of the host galaxies, which agrees with the
results reported in \citet{ode96}, \cite{bar08}, and
\citet{mar09}. The bar fraction rises from early- to late-type disks,
or in other words, from bulge-dominated galaxies to disk-dominated
galaxies (Figs. \ref{morph}a and \ref{eff}a). This appears to indicate
that bars in bulge-dominated disks are more likely to be
dissolved. However, the processes responsible cannot have been
important since $z=0.8$, because the difference in the bar fraction
between bulge-dominated galaxies and disk-dominated galaxies
($\sim25\%$, Fig. \ref{morph}) is much larger than the decline in the
bar fraction with redshift ($8\%$, Table \ref{t_evol}). This is also
illustrated in Fig. \ref{morph_z}, where we plot the bar fraction as
a function of Hubble type for a low and high redshift subsample. The
figure shows that the bar fraction in bulge-dominated galaxies was
already low at higher redshifts and implies that not many bars in
these galaxies could have been dissolved since $z>0.60$. In addition,
we observe that all morphological types contribute to the decline in
the bar fraction by $8\%$ from higher to lower redshifts (see Table
\ref{t_evol}).

Figure \ref{morph_z} seems to indicate that the $f_{bar}$-morphology
relation does not significantly change with redshift. This would
suggest that bars are typically formed or
destroyed during processes in which the morphology of the disk is
emerging or changing. In other words, bars are not dissolved in, for
instance, S0 galaxies, but can be destroyed during the processes in
which a disk galaxy is transformed into a S0. On the other hand, a
scenario in which bars are constantly formed and dissolved without
altering the relative fractions across the Hubble sequence would also
be consistent with our data. However, this would require a high degree
of fine tuning between formation and destruction processes and is
rather unlikely.

From a theoretical perspective, there are several studies indicating
that present-day bars are rather robust and not easily destroyed
\citep{she04,ath05,mar06,deb06}. Typically, a massive central mass
concentration would be required to dissolve a bar. However,
present-day super massive black holes, central dense stellar
clusters, or the inner parts of bulges are not massive enough to
affect bars significantly. However, it has been argued that
a central mass concentration with the effects of gas inflows
can lead to bar dissolution \citep{bou05}. In this picture, bars would
become weaker and weaker and start to resemble lenses, which are
preferentially found in early-type disks \citep{kor79,kor04} and can
be interpreted as dissolving bars. In this context, it is interesting
to note that we find, on average, weaker bars in early-type disks
compared to late-type disks (Fig. \ref{barmor}, lower
panel). However, our results appear to be consistent with a scenario
in which bars are rather stable and long-lived, but the possibility
that bar destruction and reformation also plays a crucial role cannot
be ruled out.

\subsection{The distribution of barred galaxies in clusters}
In Sect. \ref{resul}, we have shown that the bar fractions in clusters
and in the field are essentially the same. This result indicates that
clusters neither significantly foster nor prevent the formation of
bars and if there are processes leading to the destruction of bars in
clusters, they also act in the field. We point out that these
  results are based on relatively small samples and a confirmation
  with a larger sample would be desirable. Our findings are in
agreement with the study by \citet{van02}, who investigated the bar
fraction in field, group, and cluster environments. Considering
  the galaxy distribution in clusters, we have found
that the bar fraction in the cluster center is higher than its average 
value (Fig. \ref{cdist}). Similar results were reported for local clusters
\citep{tho81,and96}. These findings indicate that the specific
conditions in cluster centers support bar formation or help to avoid
bar destruction. The large relative velocities of the galaxies in
cluster centers prevent galaxy mergers, but galaxy flybys and the
corresponding interactions are frequent. There is theoretical
\citep{ger90,mih94,nog96} and observational \citep{elm90,giu93,var04}
evidence that galaxy-galaxy interactions trigger bar formation.

The triggered formation of bars in the cluster center could also be
responsible for the size distribution of bars. Most galaxies hosting
bars with sizes larger than the mean bar size for cluster
galaxies (3.68 kpc) are located within $R_{CC}/R_{200}<0.5$ (Fig.
\ref{blcc200}). Although the number of objects studied is small, this
is a remarkable result. Is it possible to relate the presence of these
large bars at the cluster centers to the interactions proposed to
instigate bar formation? One plausible scenario would be that the
formation of these large bars was triggered in the cluster center
and that the bars are therefore rather young. The fact that most of
the host galaxies of these bars are of intermediate Hubble type appears
to support this picture. Simulations of disk
galaxies indicated that bars grow rapidly just after their
formation \citep{ber06} and become significantly shorter once the
buckling instability occurs \citep{mar04}. Conversely, large
bar sizes can also be interpreted as signs of maturity
\citep{deb00,val03} or just as an intermediate stage during the
evolution of a barred disk galaxy \citep{cur06}. However, it is
important to point out that the disk galaxies simulated in theoretical 
studies of bar formation and evolution are typically isolated or part
of a cosmological simulation and the specific processes in clusters
are not considered.

\section{Summary and conclusions}\label{sum}
We have searched for bars in 945 galaxies, drawn from a parent sample
of 1906 disk galaxies from the EDisCS project. We used $HST/ACS$
images taken in the $F814W$ filter and restricted our sample to
$I_{auto}<23$ mag. The selection of disk galaxies (S0--Sm) was based
on their visual classifications. We identified and characterized bars
based on ellipse fits to the surface brightness distribution and
quantitative criteria. After excluding unsatisfactory fits and highly
inclined systems ($>60^{\circ}$), our bar analysis was performed for
the remaining 945 disk galaxies. Spectroscopic observations were
available for a subsample of 238 galaxies. Based on the corresponding
redshifts and cluster assignments, we evaluated the distribution of
barred disks in 11 galaxy clusters and 4 groups in the redshift range
$0.4<z\leq0.8$ and analyzed the properties of their bars. Our main
results were:
\begin{enumerate}
\item
The total optical bar fraction, averaged over the entire sample
covering the redshift range $z=0.4-0.8$, is $25\%$ ($20\%$ for strong
bars, i.e., bar ellipticity $>0.4$). This is in good agreement with
earlier studies at intermediate redshifts. The corresponding bar
fractions for the spectroscopically based cluster and field samples
are $24\%$ and $29\%$, respectively. Hence, the occurrence of bars in
clusters is roughly the same as in the field.
\item
We find that the bar fraction increases towards later Hubble types 
(form $\sim10\%$ for S0 to $>30\%$ for Sc). Interpreting this relation
as being due to the decreasing prominence of the bulge, our result is 
in close agreement with the findings of \citet{bar08}. It suggests 
that the size (or mass) of the bulge has an impact on the probability
that a bar forms or survives in a disk galaxy. The relation does not
change with redshift, indicating that bars form or dissolve only when 
the disk changes its morphology.
\item
The bar fraction as a function of effective radius exhibits a striking 
increase (from $\sim15\%$ to $\sim45\%$) towards larger radii. This
result is expected in view of the morphology-bar fraction relation
(point 2), but is much more pronounced than the latter. This indicates
that it is really the structure of the disk that strongly affects bar
formation and survival.
\item
The bar size and bar ellipticity distributions of our sample galaxies
are similar to those of other studies. The majority of bars
have sizes $<5$ kpc, as expected for un-deprojected samples. The
average bar size is rather constant along the Hubble sequence
($a_{bar}=2.5-3.5$ kpc), while the average bar ellipticity
increases towards later Hubble types (from $e_{bar}<0.4$ for S0 to
$e_{bar}>0.6$ for Sd/Sm/Im). We suspect that this result is strongly
affected by the fact that large and luminous bulges cause the
ellipticities in the bar region to be lower.
\item
We find a somewhat higher bar fraction ($\sim31\%$) close to the
centers of the clusters than at larger clustercentric distances
($\sim18\%$). This is consistent with earlier results for the Virgo
and Coma clusters. Moreover, bars in clusters are on average longer
than in the field and preferentially located close to the cluster
center. Most bars with sizes above the average ($a_{bar}=3.68$ kpc)
are located at $R_{CC}/R_{200}<0.5$. We have to point out that these
results are based on the relatively small sample of spectroscopically
confirmed cluster members and might therefore be affected by small
number statistics.
\end{enumerate}
It is interesting that we have not found a significant difference in
either bar fractions or properties between cluster and field galaxies,
but that the distributions of barred galaxies and bar sizes inside
clusters appears to be a function of clustercentric distance. The
disk galaxy properties related to bars do not differ fundamentally
between cluster and field environments, but the specific conditions in
cluster centers appear to be favorable for bar formation and survival,
and even for the creation of relatively long bars.

\begin{acknowledgements}
This paper is based on observations collected at the European Southern
Observatory, Chile, as part of large program 166.A--0162 (the ESO
Distant Cluster Survey). The Dark Cosmology Centre is funded by the
Danish National Research Foundation.
\end{acknowledgements}

\end{document}